\documentclass[twocolumn,aps,prb
groupedaddress,nofootinbib,amsmath,amssymb
]{revtex4-2}

\usepackage{tikz}

\usepackage{braket}

\usepackage[colorlinks=true,
    linkcolor=blue,
    citecolor=red,
    filecolor=magenta,      
    urlcolor=cyan,
    pdftitle={Markov},]{hyperref}


\begin{document}
\title{The ballistic to diffusive crossover in a weakly-interacting Fermi
gas}

\author{Jerome Lloyd}
\affiliation{Department of Theoretical Physics, University of Geneva, 24 rue du G{\'e}n{\'e}ral-Dufour, 1211 Gen{\`e}ve 4, Switzerland}

\author{Tibor Rakovszky}
\affiliation{Department of Physics, Stanford University, Stanford, CA 94305, USA}

\author{Frank Pollmann}
\affiliation{Department of Physics, T42, Technische Universit{\"a}t M{\"u}nchen, James-Franck-Stra{\ss}e 1, D-85748 Garching, Germany}
\affiliation{Munich Center for Quantum Science and Technology (MCQST), Schellingstr. 4, D-80799 M\"unchen, Germany } 

\author{Curt von Keyserlingk}
\affiliation{Department of Physics, King's College London, Strand WC2R 2LS, UK}

\begin{abstract}
Charge and energy are expected to diffuse in interacting systems of fermions at finite temperatures, even in the absence of disorder, with the interactions inducing a crossover from the coherent and ballistic streaming of quasi-particles at early times, to incoherent diffusive behavior at late times. The relevant crossover timescales and the transport coefficients are both controlled by the strength of interactions. In this work we develop
a numerical method to simulate such systems at high temperatures, applicable in a wide range of interaction strengths, by adapting Dissipation-assisted Operator Evolution (DAOE) to fermions. Our fermion DAOE, which approximates the exact dynamics by systematically discarding information from
high $n$-point functions, is tailored to capture non-interacting dynamics exactly, thus providing a good starting point for the weakly interacting problem. Applying our method to a microscopic model of weakly interacting fermions, we numerically demonstrate that the crossover from ballistic to diffusive transport happens 
 at a time $t_D\sim1/\Delta^{2}$ and that the diffusion constant  similarly scales as $D \sim 1/\Delta^2$,
where $\Delta$ is the interaction strength. We substantiate this scaling with a Fermi's golden rule calculation in the operator spreading picture, interpreting $t_D$ as the fermion-fermion scattering time and lifetime of the single-particle Green's function.

\end{abstract}
\maketitle

\section{Introduction}


Understanding the transport properties of materials is a central task in physics, providing the essential link between experiment and theory in most condensed matter systems. The theoretical study of transport is also deeply connected to fundamental ideas in quantum chaos and many-body thermalization~\cite{RigolReview}. More recently, somewhat less obvious but nevertheless powerful connections have been drawn between transport and the study of entanglement growth, scrambling and information spreading in many-body body systems ~\cite{rakovszky2018diffusive,khemani2018operator,parker2019universal,lucas2019operator}.

One unexpected result of the above connections has been to
inspire new approaches for calculating transport properties numerically. Simulating quantum transport
on classical computers \emph{exactly} is a difficult task because
it requires simulating real-time dynamics to late times,
and the memory required to store a time evolved state tends to grow
exponentially in time. On the other hand, in generic systems one expects that universal transport features emerge at late times which are captured by hydrodynamics, due to the system locally approaching thermal equilibrium~\cite{forster2018hydrodynamic}. Improved understanding of the generic
behaviour of operators under Heisenberg evolution has inspired new approximation schemes aimed at bridging these two levels of description in order to calculate transport coefficients from microscopic models with high accuracy~\cite{parker2019universal,rakovszky2022dissipation,klein2022time,artiaco2023efficient,white2023effective,Keyserlingk2022operator}. 


So far these methods have focused on systems of lattice spins, such as Ising or Heisenberg spin chains. However, given their ubiquity, it is perhaps more important to develop numerical methods for fermionic systems in order to describe the transport properties of arguably \emph{the} archetypal condensed matter system, namely the interacting electron gas. Such methods could be of use in taming outstanding experimental mysteries relating to the transport properties of strongly-correlated electron systems (e.g., \cite{strange_metals_sean}), but also help in quantitatively reconciling effective theories with experiment and predicting trends in transport properties with changes in experimental conditions. They would also be of use in guiding and benchmarking transport experiments in analogue quantum simulators \cite{noneq_fermi_hubbard_ott,noneq_fermi_hubbard_monika,strange_metal_cold_atom1,strange_metal_cold_atom2}.

 In this paper we generalize one such method to fermions, and use it to probe the physics of an interacting fermion system. The method, called Dissipation-assisted Operator Evolution (DAOE) evolves observables in real time, using the Heisenberg picture, and applies a truncation scheme based on the size of their spatial support. The intuition behind this approximation, substantiated by numerical and analytical calculations in Refs. \onlinecite{rakovszky2022dissipation,Keyserlingk2022operator}, is that the contribution to transport coefficients from highly non-local observables decreases sharply (indeed, exponentially~\cite{Keyserlingk2022operator}) with their support. 

We expect that similar approximations should apply to systems of interacting fermions. Indeed, the famous BBGKY hierarchy and the related Boltzmann equation \cite{kardar_book_particles} are similar in spirit, truncating high order fermionic correlations to achieve a closed set of equations for the low order ones. However, the original formulation of DAOE~\cite{rakovszky2022dissipation} does not directly capture this intuition, as it is formulated in terms of spin variables. While, at least in one dimension (which is where DAOE currently applies), the two can be related by the Jordan-Wigner transformation, the relevant notion of ``operator support'' is very different in the two cases, due to the non-locality of the transformation; indeed, 2-body fermionic operators map onto long non-local strings in the spin language. As a consequence, applying DAOE in the naive way fails to capture even the exactly solvable free fermionic case, as we demonstrate explicitly below. Instead, we develop a modified approximation scheme, named fermionic DAOE (fDAOE), which keeps few-body fermionic observables exactly, while truncating higher order ones in a tunable way (tuned by the strength of an artificial dissipation). It is therefore exact for free fermions and provides a good starting point to incorporate the effects of electron-electron scattering. 

A particularly interesting test case for our fermionic method is in the limit of weak, but non-vanishing interactions. In the fully non-interacting limit, and in the absence of disorder, electrons propagate \emph{ballistically} and coherently (i.e., the fermion propagator exhibits a characteristic interference pattern due to quantum mechanical effects). However, when many electrons interact, their collisions tend to wash out interference effects at long times, typically leading to a \emph{diffusive} charge transport at finite temperatures~\cite{ashcroft2022solid}. One therefore expects a ballistic-to-diffusive crossover which happens at parametrically late times when interactions are weak (correspondingly, the resulting diffusion constant is also parametrically large in this regime). These characteristically large time scales make it challenging to capture the behavior of these systems accurately. Novel numerical approaches are therefore needed. The fact that fDAOE becomes exact when interactions are turned off makes it a promising candidate to tackle this challenge. 

In the weakly interacting regime, it is possible to use perturbation theory (in the guise of Fermi's Golden Rule, as we will discuss in Sec.~\ref{sec:FGR_op_spread}) to argue that diffusion constants, and scattering timescales go as $O(1/\Delta^2)$ where  $\Delta$ is the interaction strength. However, such arguments are ultimately uncontrolled, as witnessed by the fact that weakly interacting integrable models have interactions but do not exhibit diffusion at small $\Delta$. So, while FGR can be used to estimate the timescales when a free fermion approximation breaks down\footnote{Even this estimate is quite delicate, and difficult to make rigorous, at least in one dimension, as we discuss in Sec.~\ref{sec:FGR_op_spread}.}, it does not give access to transport features at asymptotically late times, which are inherently non-perturbative. 
This lack of analytical control makes necessary the development of the numerical tools in this work. 

In this paper, we consider a non-integrable one-dimensional model of fermions with a tunable interaction strength $\Delta$. We demonstrate the power of fermionic DAOE by showing that it can obtain well-converged results for a whole range of interactions, from free to strongly coupled fermions, unlike either brute-force MPS methods~\cite{schollwock2011density,paeckel2019time}, which fails to capture the interacting case due to the rapid growth of entanglement, or spin-based DAOE, which fails already at the non-interacting level. By following the dynamics to late times, we find the expected crossover to diffusive transport, and fit its time scale to be $t_D \propto 1/\Delta^2$. We also fit the diffusion constant itself and find evidence of similar scaling with $\Delta$. Finally we outline a Fermi's golden rule perturbative calculation, which supports this scaling, and provide a picture for the ballistic to diffusive crossover in the language of operator spreading.

\section{DAOE method}

DAOE~\cite{rakovszky2022dissipation} is an algorithm for calculating dynamical
correlation functions in interacting quantum systems. It can be used to extract transport coefficients via linear response theory. In this paper we focus on calculation of the diffusion constants, which measures how initial perturbations in globally conserved densities (charge, spin etc.) spread throughout the system. We first describe how this is done in the existing version of DAOE (which we will refer to as ``spin DAOE''), before describing its modification in the fermionic context.

\subsection{Spin DAOE}

Denote by $q_{x}$ the local density at site $x$ of the globally conserved quantity $Q$. We consider one-dimensional systems of length $L$, and take $x=0$ to label the site in the centre of the chain. Without loss of generality assume that $\left\langle q_{x}\right\rangle =0$
at equilibrium, and normalised such that $\sum_{x}\left\langle q_{0}q_{x}\right\rangle =1$, where $\langle A \rangle = \text{tr}(A\rho)$ and $\rho$ is the equilibrium density matrix. We work at infinite temperature throughout, so that $\langle A\rangle=\text{tr}(A)/\mathcal{N}$, $\mathcal{N}$ being the Hilbert space dimension. The associated diffusion constant is then given by $D\equiv\lim_{t\rightarrow\infty} D(t)$ where 
\begin{equation}
    D(t)=\lim_{L\rightarrow\infty} \partial_t \sum_{x}\frac{x^{2}}{2}\langle q_{x}|\mathcal{U}(t,0)|q_{0}\rangle.
\label{eq:D_1st}
\end{equation}
Here the superoperator $\mathcal{U}$ evolves operators in the Heisenberg picture,
\begin{equation}
\langle A|\mathcal{U}(t_{2},t_{1})\equiv\langle U^{\dagger}(t_{2},t_{1})AU(t_{2},t_{1})|,
\end{equation}
 and we introduced an inner product between operators $\langle A|B\rangle\equiv\bigl\langle A^{\dagger}B\bigr\rangle$.
 
 The difficulty of evaluating Eq.~\eqref{eq:D_1st} lies in the fact that in generic systems, the memory required to accurately store the time-evolved operator, for examples in a matrix product operator (MPO) representation, grows exponentially in time, due to the linear Lieb-Robinson light cone, which makes an exact evaluation of the necessary correlation functions prohibitively expensive \cite{prosen2007efficiency,jonay2018coarse}. In DAOE, this is circumvented by truncating the operator content, on the premise that much of the information stored in non-local operators is irrelevant for computing the overlaps $\langle q_x(t) q_0\rangle $ between local observables, as elaborated below.

The algorithm is implemented as follows. In Eq.~\eqref{eq:D_1st},
we replace the unitary evolution $\mathcal{U}(t,0)$ with the periodically dissipated evolution $\grave{\mathcal{U}}(t,0)\equiv\prod_{j=0}^{t/T-1}[\mathcal{U}(t_{j+1},t_{j})\mathcal{G}_{\ell_{*},\gamma}]$.  The dissipation superoperator $\mathcal{G}_{\ell_{*},\gamma}$ is applied with a period $T$ (with $t_{j}=jT$ and $t/T$ assumed to be integer) and designed to suppress operators larger than some cutoff length $\ell_{*}$. In a spin-1/2 system, $\mathcal{G}_{\ell_{*},\gamma}$ can be defined by its action on \emph{Pauli strings}   $|\sigma^{\mu}\rangle$. These are tensor products of matrices $I,\sigma^{x,y,z}$ on the different sites of the chain, which form a basis (of size $4^L$)
for the space of all operators acting on the spin chain. The dissipator
then takes the form 
\begin{equation}
\mathcal{G}_{\ell_{*},\gamma}|\sigma^{\mu}\rangle=e^{-\gamma\max(0,\ell(\mu)-\ell_{*})}|\sigma^{\mu}\rangle,\label{eq:DAOEdefn}
\end{equation}
 where $\ell(\mu)$ -{}- the `length' of operator $\sigma^{\mu}$
(also known as the \emph{Pauli weight}) -{}- is the number of non-identity
matrices in the string $\sigma^{\mu}$. Using this nomenclature, note
that the operator $\sigma_{1}^{z}\sigma_{27}^{x}$ has length $\ell=2$,
even though it is spread over a large spatial region (of diameter
$27$). This is the form of DAOE that appeared in Refs. \cite{rakovszky2022dissipation,Keyserlingk2022operator}.

How does this artificial dissipation help in evaluating Eq.~\eqref{eq:D_1st}?
As was shown in Ref. \cite{rakovszky2022dissipation}, the dissipation cuts off the growth
of the effective entanglement of the state $|q_{x}(t)\rangle$
(the operator space entanglement entropy, or OSEE~\cite{prosen2007operator,prosen2007efficiency}). This means that $|q_{x}(t)\rangle$ can be efficiently represented as a \emph{matrix product state} (MPS), which allows one to use standard tensor network techniques, such
as Time-Evolving Block Decimation (TEBD)~\cite{vidal2003efficient} to perform the time evolution
efficiently; the errors made by truncating to a finite bond dimension
will depend only on the amount of dissipation used, and they will
remain bounded at all times, unlike the case for purely unitary dynamics where entanglement grows without bound~\cite{prosen2007efficiency,jonay2018coarse}.
The operator $\ensuremath{\mathcal{G}_{\ell_{*},\gamma}}$ defined by Eq.~\eqref{eq:DAOEdefn} itself has a compact representation as a matrix product operator (MPO) representation with bond dimension $\ell_{*}+1$,
which makes its application efficient. 

While the introduction of the dissipation renders the numerical calculation
of correlations feasible, it will also introduce errors, i.e.~the
diffusion constant of the dissipative dynamics, $D_{\ell_{*},\gamma}$
will be different from the true physical diffusion constant $D$. The insight behind DAOE is that the error made in this approximation should be well controlled when the dissipation is sufficiently weak, allowing for an accurate estimation of the true diffusion constant. The reason behind this is that once an operator becomes highly non-local, its contribution to correlation functions of local observables becomes suppressed, due to a combination of entropic and interference effects~\cite{Keyserlingk2022operator}. 
More quantitatively, in Ref. \cite{Keyserlingk2022operator} some of us conjectured, based on numerical evidence
and analytical arguments, that this error is suppressed exponentially,
as $|D-D_{\ell_{*},\gamma}|\leq ae^{-b\ell_{*}}$ at large $\ell_{*}$.
More practically, one calculates $D_{\ell_{*},\gamma}$ for
decreasing values of $\gamma$ until the limit of the available numerical
resources is reached\footnote{The weaker the dissipation, the larger the maximum of the OSEE reached
during the dynamics, implying that one will need increasingly large
bond dimensions to get accurate results.}, and then uses the results to perform an extrapolation back to the
non-dissipative case: $D=\lim_{\gamma\to0}D_{\ell_{*},\gamma}$. This limit should be independent of the choice of $\ell_*$, which provides a further consistency check on the results~\cite{rakovszky2022dissipation}.

\subsection{Modified DAOE for fermions (\lowercase{f}DAOE)}

\begin{figure}
    \centering
    \includegraphics[width=\columnwidth]{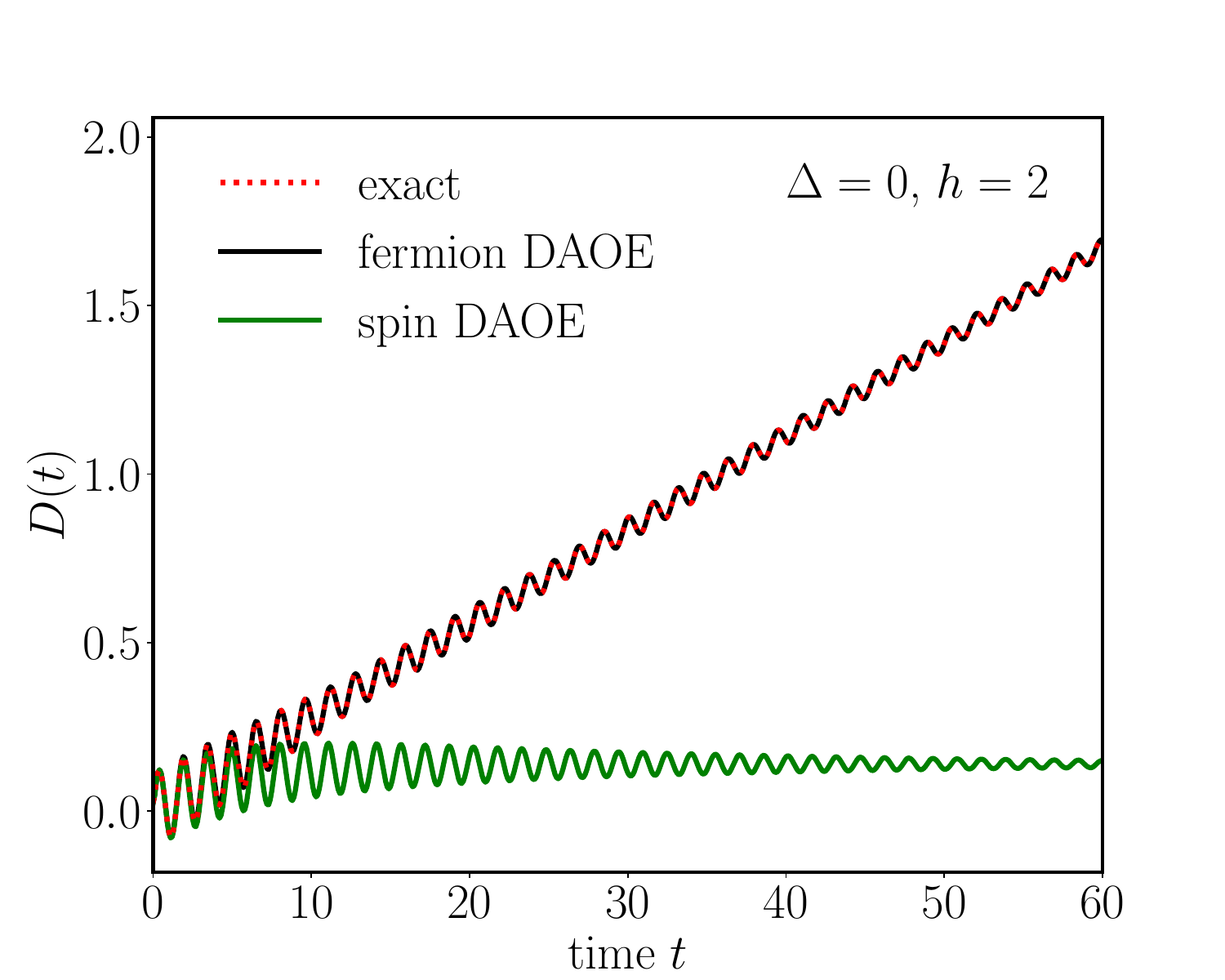}
    \caption{For the non-interacting case $\Delta=0$, $h=0$ of the model defined in \eqref{eq:H_fermions}, we show $D(t)$ vs.~$t$ as determined by different methods. The red dotted line is the exact result calculated from the free-fermion propagator $K(t)$, with the ballistic scaling $D(t) \sim t$. Fermion DAOE agrees with the exact result provided $\ell_*\geq 2$. Spin DAOE, however, gives incorrect diffusive behaviour past short times even in the $\Delta=0$ case, due to heavily truncated Jordan-Wigner strings. The DAOE parameters were taken as $\gamma=0.1$, $\ell_*=2$, $\delta t = 0.5$ in both cases (the fDAOE results are independent of the parameters in the non-interacting case, provided $\ell_* \geq 2$).}
    \label{fig:naive_DAOE}
\end{figure}

The spin DAOE we just described is formulated for systems of lattice spins, appropriate for simulating transport in quantum magnets~\cite{hess2007heat,hirobe2017one,bertini2021finite}. However, one often wants to obtain information about transport in systems of fermions, where we can ask about the transport of electric charge. The physical intuition behind DAOE is general enough that it should apply to this case as well, assuming that one is at a finite temperature where electron scattering should render transport incoherent. Thus, it is natural to formulate a fermionic version of DAOE, which is what we obtain here, working still in one dimension.

We consider Hamiltonian dynamics which conserve the total fermion charge, $[H, \sum_x n_x]=0$, with $n_x=f^\dagger_x f_x$ the local charge density and $f^\dagger_x, f_x$ the standard spin-less fermion creation-annihilation operators. To calculate charge transport, we need to evaluate connected correlations of the form $\braket{n_x(t)n_y(t')}^c$. In one dimensional systems, one possibility is to make use of the Jordan-Wigner transformation to rewrite fermionic operators in terms of spins, and then apply spin DAOE in this new formulation. However, this is complicated by the fact that the transformation between the two variables is non-local; thus, the notion of Pauli weight that we used to truncate non-local observables in spin DAOE is not the appropriate measure of locality. 

The problems with this naive approach can be observed already at the level of non-interacting fermions. In this case the time evolution of a single Fermion operator $f^\dagger_x$ maps into a coherent superposition of 1-body operators 
\begin{align}
f_{x}^{\dagger} & \rightarrow\sum_{y}K_{y,x}(t)f_{y}^{\dagger}\label{eq:free_fermion_operator_evolution}
\end{align}
where $K$ is the non-interacting lattice Feynman propagator. The time-evolved charge density takes the form $n_x(t)=\sum_{y,y'}K_{xy}(t)K_{xy'}^{*}(t)f_{y}^{\dagger}f_{y'}$. For translationally invariant Hamiltonians on the lattice, the free Feynman propagator $K_{xy}(t)$ is generically nonvanishing only within a light-cone $|y-x|<v t$, where it has amplitude roughly $1/\sqrt{t}$. The result of this is that density spreads ballistically i.e., $\braket{n_x(t)n_y(t')}^c$ is non-vanishing only when $|x-y|<vt$ (and has typical size $\sim 1/t$ within the light-cone). By Eq.~\eqref{eq:D_1st}, this implies that the time-dependent diffusion constant grows linearly as $D(t) \sim t$. 

To apply spin DAOE to this problem, we can make use of the Jordan-Wigner transformation
\begin{equation}\label{eq:JW}
f_{x}^{\dagger}=\bigl[\prod_{s<x}\sigma_{s}^{z}\bigr]\sigma_{x}^{+}.
\end{equation}
This turns the operators$f^\dagger_{y}f_{y'}$, appearing in the calculation of $n_x(t)$ into strings of spin variables stretching from $y$ to $y'$. As we say, the typical terms contributing to $n_x(t)$ have $|y-y'|=O(t)$. Therefore at sufficiently late times compared to $\ell_*$, $n_x(t)$ contains long strings of spin operators which will be subject to the DAOE dissipation under the existing scheme. This suggests that the existing DAOE approach will fail to correctly capture the ballistic spread of charge even in the non-interacting limit.

This expectation is confirmed in Fig.~\ref{fig:naive_DAOE} where we show the measured time-dependent diffusion constant of the charge density for free-fermion dynamics ($h=0,\ \Delta = 0$ in the model defined in Eq.~\eqref{eq:H_fermions} below), with DAOE truncating according to Pauli weight of the Jordan-Wigner transformed operators (green line). The results show a very rapid onset of diffusion, whereas the true behaviour, as calculated from the non-interacting propagator $K(t)$, is ballistic (red dotted line). 
%


\begin{figure*}
    \centering
    \includegraphics[width=\textwidth]{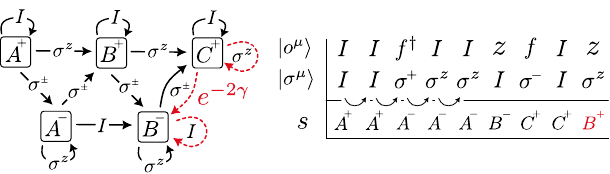}
    \caption{\emph{Left}: a 5-state finite automata for fDAOE with $\ell_*=4$. See text for details. \emph{Right}: example fermion string $\ket{f^\mu}$, corresponding Jordan-Wigner mapped Pauli string $\ket{\sigma^\mu}$, and automata trajectory $s$.}
    \label{fig:automata}
\end{figure*}

The issue lies in the use of the Pauli weight for truncating fermionic systems: this suggests a natural generalisation of DAOE that instead  truncates operators based on their fermion operator weight (`Fermi weight'). To make this notion precise, we introduce the Fermi strings $\ket{o^\mu}$, in analogy to the Pauli strings $\ket{\sigma^\mu}$. A complete on-site basis of operators is given by the four operators $o^{1,2,3,4} = I, f^\dagger, f, [f^\dagger, f]$, which have Fermi weights $0,1,1,2$ respectively. In the following we use the variable $z$ as a short-hand for $[f^\dagger, f]=2f^\dagger f - 1\!\!1$. The Fermi strings $o^{\mu}\equiv \otimes_r o_r^{\mu_r}$ are tensor products of the local basis operators on $L$ sites, and we define the Fermi weight of $o^{\mu}$ as the sum of its on-site weights. For example, the string $f_2f^\dagger_6f_6$ has Fermi weight $\ell=3$. Now, quadratic operators such as $f_{y}^{\dagger}f_{y'}(t)$ evolved under non-interacting dynamics have weight $\ell = 2$ for all times, even if $y,y'$ are well separated. The only change needed in the definition of the DAOE dissipator (Eq.~\ref{eq:DAOEdefn}) is replacing the Pauli weight with the Fermi weight. 

Provided we choose cutoff $\ell_{*}\geq2$, dissipation will not interfere with the early-time ballistic behavior of correlation functions. Comparing $D(t)$ calculated using fermion DAOE in Fig.~\ref{fig:naive_DAOE} (black line), we observe perfect agreement with the exact result. We refer to the Fermi weight truncated DAOE as fDAOE. In order for fDAOE to remain efficient, we still need to be able to write the dissipator in a compact MPO format. This turns out to be possible, following the method of \cite{crosswhite2008finite}, as we now describe.

We first design a `finite state automata' $\mathcal{W}$ that takes as its input a Fermi string $\ket{o^\mu}$ (equivalently a Pauli string $\ket{\sigma^\mu}$, as the two can be mapped via the Jordan-Wigner transform), calculates the corresponding Fermi weight $\ell(\mu)$, and outputs a weighted copy $e^{-\gamma\text{max}(0, \ell(\mu)-\ell_*)} \ket{o^\mu}$. The automata $\mathcal{W}$ is a machine that moves along the input string $\ket{o^\mu}$, reading the operator at each position $x$ from left to right, and updating its internal state based on this new information and its current state, finally outputting the desired the desired weight $e^{-\gamma\text{max}(0, \ell(\mu)-\ell_*)}$. Having constructed our finite state automata, we finally have to convert it into an MPO. This can be done with virtual bond dimension $\ell_*+1$. We explain the case $\ell_*=4$ in detail, with the extension to other $\ell_*$ straightforward. We restrict ourselves the case where all Fermi strings have even parity, which provides no loss of generality when the system obeys charge conservation. 

The fDAOE automata for $\ell_*=4$ is drawn to the left of Fig.~\ref{fig:automata}. We consider the automata's action on the example fermion string given in the top row of the table on the right of Fig.~\ref{fig:automata}. The fermion string is mapped to the Pauli string in the second row via the Jordan-Wigner transform (this step is required as numerically our MPO must operate on Pauli strings). The machine starts at the first site of the string in an initial state $s = A^+$. It then reads the operator at the current site. Following the map on the left of Fig.~\ref{fig:automata}, there are then three possibilities: if the operator is $I$, the automata remains in state $A^+$; if the operator is $\sigma^z$, the automata updates its state to $s=B^+$; otherwise, the operator is either $\sigma^+$ or $\sigma^-$, and the automata sets its new state as $s=A^-$. It then moves to the second site of the chain, and repeats the above steps, following the new directions based on its current state. The state label allows to keep track of the Fermi weight and whether $\ell_*$ has been exceeded: for every red dashed line the automata follows (e.g.~from state $B^-$ to $C^+$), the input string is multiplied by a damping factor $e^{-2\gamma}$. The resulting automata `trajectory' for the example string above is given in the third row of the table in Fig.~\ref{fig:automata}: the string $\ket{o^\mu}$ has Fermi weight $\ell = 6$ and is correctly weighted by the automata with a factor $e^{-2\gamma}$. 

The key point is that to correctly account for the non-locality of the Jordan-Wigner transform, the automata must keep track of the fermion parity i.e.~the number of fermions to the left of its current position.
A pair of spatially separated fermions $f^\dagger, f$ (connected by a string of identity operators) maps to a pair of $\sigma^\pm$ operators connected by a string of $\sigma^z$ operators in the Pauli basis. In order for these $\sigma^z$ to not count toward the Fermi weight, the automata must switch to the negative parity branch $A^-,B^-,\ldots$ where the roles of $\sigma^z$ and $I$ are exchanged. This parity switching is the reason why fDAOE correctly captures the non-interacting fermion dynamics, while spin DAOE truncates quadratic strings.  

Given the finite state machine it is then simple to construct the corresponding MPO. We refer to \cite{crosswhite2008finite} for details and simply state the result for $\ell_*=4$. The local MPO tensor $W^{ii'}_{ab}$ acts diagonally on the physical leg index $i = I,+,-,Z$, $W^{ii'}_{ab}\propto\delta_{ii'}$, and as a matrix on the virtual leg index $a = 0,\ldots,\ell_*$. Owing to the even parity constraint, $\ell_*$ must be even. Explicitly the matrices are given by

\begin{gather}\renewcommand*{\arraystretch}{.6}
    W^{II} = \left(\begin{array}{ccc|cc}
    1 & 0 & 0 & {} & {} \\
    0 & 1 & 0 & {} & {} \\
    0 & 0 & 1 & {} & {} \\\hline
    {} & {} & {} & 0 & 1 \\
    {} & {} & {} & 0 & \kappa
    \end{array}\right), \hspace{.5cm} 
    W^{ZZ} = \left(\begin{array}{ccc|cc}
    0 & 1 & 0 & {} & {} \\
    0 & 0 & 1 & {} & {} \\
    0 & 0 & \kappa & {} & {} \\\hline
    {} & {} & {} & 1 & 0 \\
    {} & {} & {} & 0 & 1 
    \end{array}\right), \nonumber \\
     W^{\pm\pm} = \left(\begin{array}{ccc|cc}
    {} & {} & {} & 1 & 0 \\
    {} & {} & {} & 0 & 1 \\
    {} & {} & {} & 0 & \kappa \\\hline
    0 & 1 & 0 & {} & {} \\
    0 & 0 & 1 & {} & {} \\
    \end{array}\right).
\end{gather}
Here we split the matrices into parity blocks $++$, $+-$, $-+$, $--$: the even parity ($+$) indices run from 0 to $\ell_*/2$, and the odd parity indices ($-$) from $\ell_*/2+1$ to $\ell_*$. The damping factor $\kappa = e^{-2\gamma}$ is applied on transitions corresponding to the red dashed lines in Fig.~\ref{fig:automata}. The MPO is contracted on the left boundary with the vector $v_L=(1,0,\ldots,0)$ and with $v_R=(1,1,\ldots,1)$ on the right.

The fDAOE dissipator can hence be applied in an efficient way to operators represented as matrix-product states $\ket{q(t}$. This is the first main result of this work. 

\section{MODEL}

For the rest of this work we focus on the physics of hopping spinless fermions with a staggered chemical potential:
\begin{widetext}
\begin{equation}
H=\underbrace{-\sum_{x}\Bigl(\frac{J}{2}(f_{x+1}^{\dagger}f_{x}+\mathrm{h.c.})+h(-1)^{x}(1-2n_x)\Bigr) }_{H_0}+\underbrace{\Delta\sum_{x}(1-2n_{x})(1-2n_{x+1})}_{\Delta\times V}.\label{eq:H_fermions}
\end{equation}

By the Jordan-Wigner mapping, the Hamiltonian can also be written as the staggered-field XXZ Hamiltonian,
\begin{equation}
H=\sum_{x}\Bigl(\frac{J}{2}\bigl(\sigma_{x}^{+}\sigma_{x+1}^{-}+\mathrm{h.c.}\bigr)+h(-1)^x\sigma_x^z\Bigr)+\Delta\sum_{x}\sigma_{x}^{\mathrm{z}}\sigma_{x+1}^{z}.\label{eq:H_spins}
\end{equation}
\end{widetext}

This model (first introduced in \cite{huang2013scaling}) is non-integrable if all three couplings, $J,h,\Delta$ are non-zero. In particular, switching on the staggered field $h$ breaks the integrability of the XXZ model. We will focus on the regime where $\Delta \ll J,h$. At $\Delta=0$ the fermions are non-interacting, and exhibit the ballistic transport we observed in Fig.~\ref{fig:naive_DAOE}. Turning on $\Delta$ turns this into a problem of interacting fermions. We will be interested in how charge transport is affected by these interactions. 

\section{Numerical results}

\begin{figure*}
    \centering
    \includegraphics[width=\textwidth]{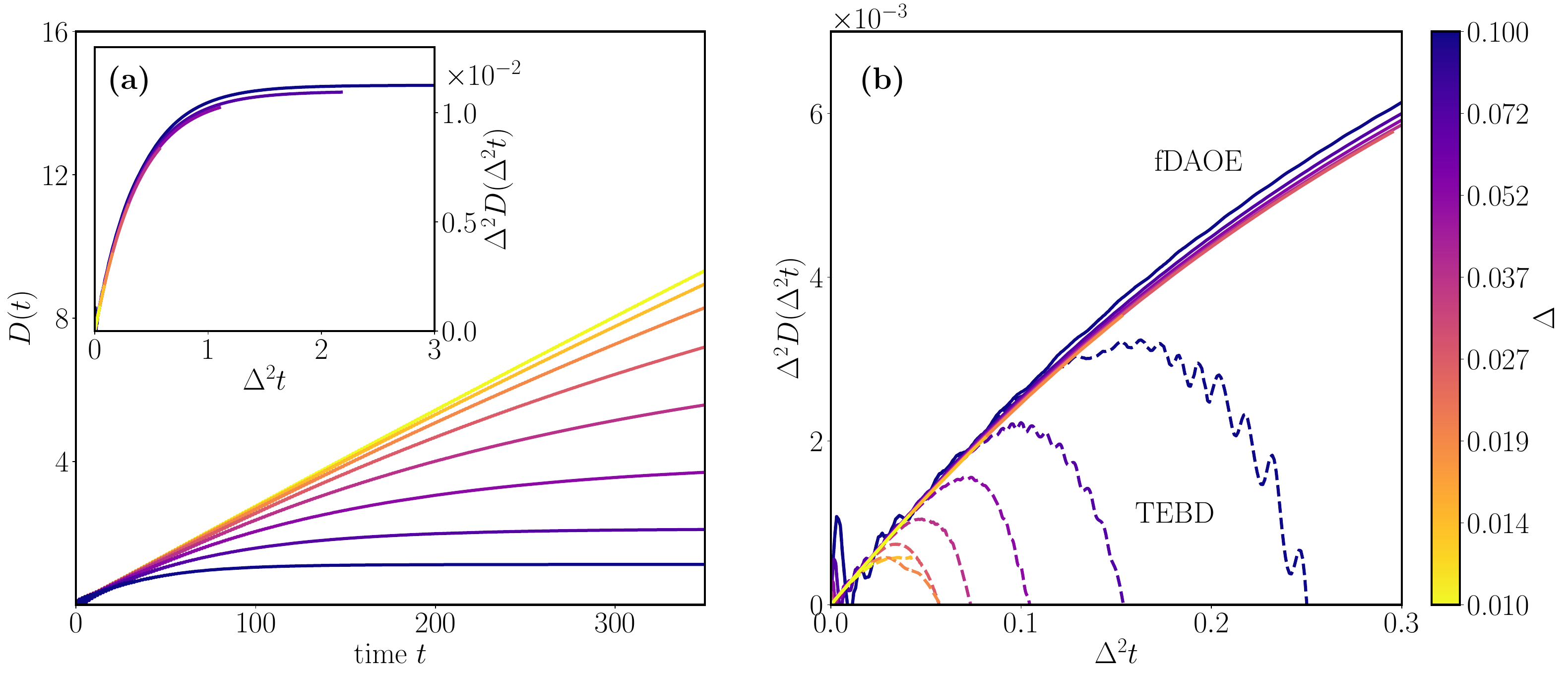}
    \caption{\textbf{(a)} Growth of time-dependent diffusion constant $D(t)$, for eight different interaction strengths $\Delta$, as marked on the right-hand colorbar. For increasing $\Delta$, the timescale at which transport becomes diffusive decreases, with a late time diffusive plateau reached for the strongest interactions around $t=200-300$.  \textbf{Inset}: scaling collapse of rescaled diffusion constant $\Delta^2D(\Delta^2 t)$. The diffusion constant grows linearly at early times $\Delta^2 t \ll 1$, before a crossover to diffusive transport occurs at late times $\Delta^2 t \gg 1$, where the curve saturates. \textbf{(b)} Zoom on $\Delta^2D(\Delta^2 t)$ for short times, with TEBD simulations (dashed lines) overlayed. For all $\Delta$, TEBD fails to reach the crossover timescale $\Delta^2 t \sim 1$, due to an accumulation of truncation errors arising from unchecked entanglement growth. For both figures, we fix DAOE parameters $\gamma=0.1$, $\ell_*=2$ and maximum bond dimension $\chi=96$ for fDAOE, $\chi=128$ for TEBD. }
    \label{fig:crossover}
\end{figure*}

\begin{figure*}
    \centering
    \includegraphics[width=\textwidth]{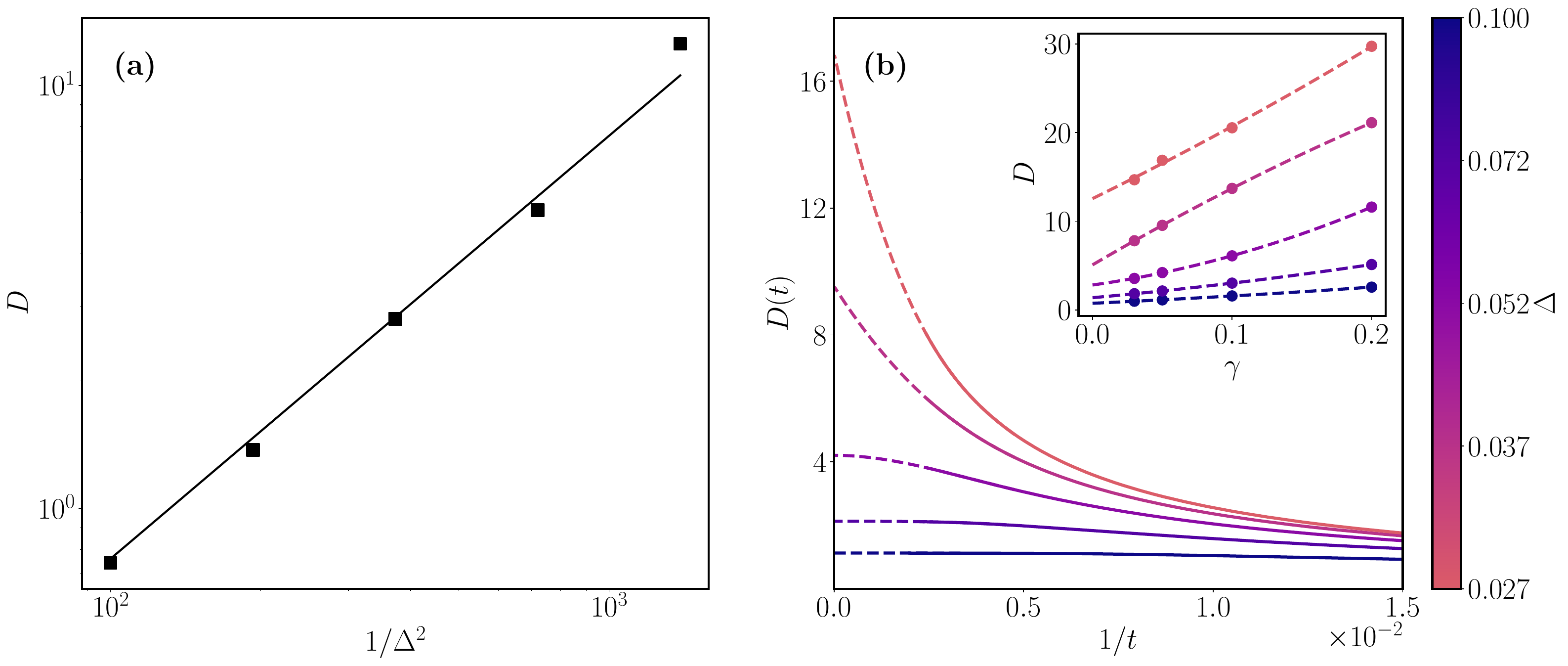}
    \caption{\textbf{(a)} Late time diffusion constant $D \equiv \lim_{t\to\infty} D(t)$ as a function of interaction strength $\Delta$ (we show only the largest five $\Delta$, as the smaller $\Delta$ data are not sufficiently diffusive at the maximum run-time to obtain accurate exrapolation). The data is extrapolated in $\gamma \to 0$. The fit is the function $D = d\Delta^{-2}$, with the fit parameter $d=0.01$. \textbf{(b)} Extrapolation of $D$: we fit $D(t)$ to a cubic polynomial in the inverse time variable $1/t$ at late times, with $D$ as the $y$-axis intercept. Solid lines are data and dashed lines are the fit. Data shown is for $\gamma = 0.05$. \textbf{Inset} Extrapolation in $\gamma$: for each data point the diffusion constant is obtained by the previous method, then we fit the data to a quadratic polynomial in powers of $\gamma$ (dashed lines).}
    \label{fig:diffusion}
\end{figure*}

We now turn to applying fDAOE to study the ballistic-to-diffusive crossover in the model of interacting fermions in \eqref{eq:H_fermions}. We fix the model parameters $J=1$ and $h=2$, while varying the interaction strength $\Delta$. As explained above, the model is no longer integrable when $\Delta \neq 0$, and diffusive behaviour is expected to arise at late times. 

Starting from the infinite temperature initial state, we use fDAOE to simulate the evolution of the density-density correlators $C(x,t) = \braket{q_x(t)|q_0(0)}$, with $q_x = \sigma^z_x$ and normalised as $\sum_x C(x,0) = 1$. Here $x=0$ denotes the central chain site. This spin correlator is related to the connected fermion charge-charge correlator by $C(x,t) = 4\langle n_x(t)n_0(0)\rangle^c$. We then calculate the time-dependent diffusion constant $D(t)$ using Eq.~\eqref{eq:D_1st}. All of our simulations are done using the TeNPy package \cite{hauschild2018efficient} which allows us to make use of the system's charge conservation property for more efficient simulations.  Throughout the rest of the paper we fix the (2\textsuperscript{nd} order) TEBD timestep $\delta t = 0.1$, DAOE truncation period $T=0.5$, and the chain length $L=300$, which is large enough that boundary effects are not observable on the times we study (the maximum times we reach are around $t=500$, using around 100h of wall time).

In Fig.~\ref{fig:crossover}(a) we plot the growth of the time dependent diffusion constant $D(t)$ against time $t$, for different $\Delta$. We fix DAOE parameters as $\gamma=0.1$, $\ell_*=2$ and show results for eight logarithmically spaced $\Delta$ in the interval $(0.01, 0.1)$. The data shown is for bond dimension $\chi=96$ --- we provide additional convergence data in Appendix \ref{app:convergence}. The crossover behaviour is evident for large $\Delta$, with an initial ballistic growth $D(t) \propto t$, followed by a late time plateau to diffusion $D(t) \sim D$ (the value of which generally depends on $\Delta$ as well as the DAOE parameters, see below). The results for the smallest $\Delta$ values remain approximately ballistic up to the largest times simulated (here, $t=350)$. 

To determine the how $D(t)$ depends on $\Delta$, we perform a rescaling of the time $t \to \Delta^2 t$ and the diffusion constant $D(t) \to \Delta^2 D(\Delta^2 t)$. In the inset to Fig.~\ref{fig:crossover}(a), we plot the rescaled dependence. We observe a clean scaling collapse of the data, consistent with the approximate scaling $D \propto \Delta^{-2}$ and a ballistic-to-diffusive crossover time (when the coherent transport of charge crosses over to diffusive transport) $t_D \sim \Delta^{-2}$. This scaling is consistent with a perturbative argument following Fermi's golden rule which we outline in Section \ref{sec:FGR_op_spread}.
The scaling collapse is not expected to be exact, as higher order corrections in perturbation theory will generally affect the late time value of $D(t)$. 

Our results provide a convincing picture of how diffusion arises from weak fermion-fermion scattering at late times. However, given the basic nature of the above result, one may wonder whether the same answer could have been obtained based on simpler evolution methods, such as `pure' TEBD algorithm without any truncation from DAOE. After all, in the free-fermion limit $\Delta = 0$, the system's entanglement remains low for all times, which is the regime in which TEBD is expected to be efficient. In Fig.~\ref{fig:crossover}(b) we test this notion by plotting $\Delta^2D(\Delta^2 t)$ as calculated from TEBD simulations against the same fDAOE data shown in Fig.~\ref{fig:crossover}(a): we find that for all $\Delta$, truncation errors accumulate significantly before the evolution reaches the crossover timescale $\Delta^2t = 1$ and TEBD fails catastrophically, with $D(t)$ nosediving towards unphysical negative values. We note that while the bond dimension used for TEBD (128, compared to 96 for fDAOE) is relatively low, to completely avoid truncation errors the required bond dimension must grow exponentially in time; the success of DAOE for simulating long times lies in the fact that the required bond dimension is bounded for all times by a value controlled by the dissipation parameters. Thus, fDAOE allows us to efficiently reach diffusive timescales that remain unavailable to more standard methods. 

To conclude this section we attempt to extrapolate the infinite time diffusion constant $D \equiv \lim_{t\to\infty} D(t)$, as a further check on the scaling result. We consider the five largest $\Delta$ values in the previous dataset (for smaller $\Delta$, the simulations do not reach far enough into the diffusive regime to achieve a reliable estimate), and fit the late time data to a cubic polynomial in the inverse time variable, i.e.~$D(t) \sim D - a/t + b/t^2 +c/t^3$. The resulting diffusion constants $D$ are then extrapolated in $\gamma \to 0$ for the four values $\gamma \in (0.03, 0.05, 0.1, 0.2)$, for each $\Delta$. The result is shown in Fig.~\ref{fig:diffusion}(a), with the scaling $D \sim \Delta^{-2}$: the black line is the fit $D = 0.01\Delta^{-2}$. In Fig.~\ref{fig:diffusion}(b) we show additional details of the extrapolation process, with the $1/t$ extrapolation for $\gamma = 0.05$ shown in the main figure and the subsequent $\gamma$ extrapolation in the inset. 

We make two additional observations from the data in Fig.~\ref{fig:diffusion}(b): firstly, $D(t)$ exhibits strong plateaus (in $1/t$) towards the final value (visible for the three largest $\Delta$). This is unexpected, as the hydrodynamic theory would predict power law tails as we discuss in Sec.~\ref{sec:FGR_op_spread}, see also ~\cite{Mukerjee_2006, michailidis2023corrections}. We do not have a satisfying explanation of this; it might be an artifact of the fDAOE dissipator suppressing hydrodynamic tails, or those tails might just be small in this system. 

Secondly, we note that the predicted diffusion constant generally \emph{increases} as the parameter $\gamma$ is increased. At first glance this is also somewhat surprising, as dissipation (such as between the system and an environment) is generally expected to lead to increased loss of coherence and faster approach to diffusion. We provide a heuristic interpretation of this behaviour as akin to a dissipation-induced `diffusive Zeno effect': when $\ell_*=2$, fDAOE acts to suppress growth of larger fermion strings, while leaving quadratic strings invariant. Is is easy to see that for $\gamma \to \infty$, interactions are completely blocked and no diffusive transport can occur. For large $\gamma$, fermions remain `confined' within the quadratic free-fermion space and charge continues to be transported ballistically for long times. We show addition results for this effect in Appendix \ref{app:zeno}.

\section{Fermi's golden rule, and an operator spreading picture for the diffusive crossover}\label{sec:FGR_op_spread}

\begin{figure}
    \centering
    \includegraphics[width=0.49\textwidth]{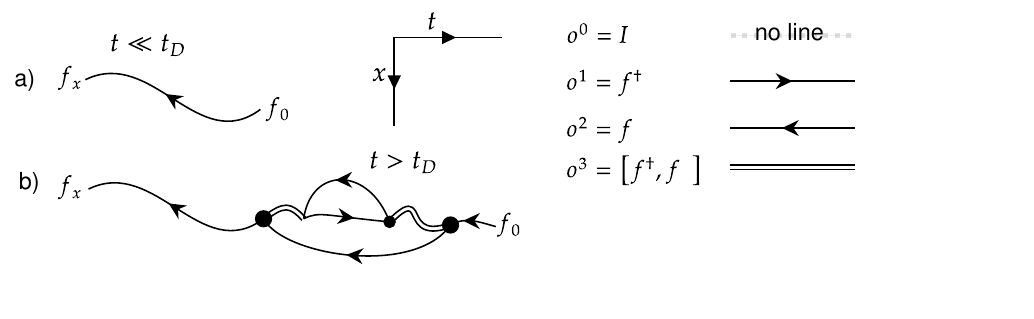}
    \caption{An operator path integral for the single particle Green's function $G(x,t)=\langle f_x(t) |f_0\rangle$ consists of a sum of operator trajectories starting at $f_x$ on the left, and evolving to $f_0$ at time $t$. (a) In the absence of interactions, the number of fermions operators is conserved, and this remains approximately true at early times when interactions are unimportant. (b) At later times, interactions cause 1-body operators to grow in Fermi weight, which leads to an (exponential in time) suppression of the amplitude $G(x,t)$.}
    \label{fig:G}
\end{figure}

\begin{figure}
    \centering
     \includegraphics[width=0.49\textwidth]{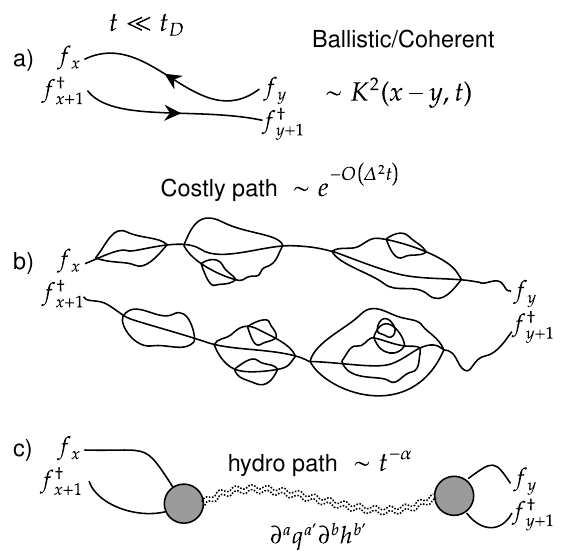}
    \caption{ A diagram of contributions to the current-current correlator. a) At early times, the dominant contribution to transport comes from paths where the fermion operators do not grow in Fermi-weight but instead ballistically spread according to the non-interacting dynamics. We identify two sorts of paths at later times. b) Those paths where the initial fermions propagate independently, but undergo forking events due to interactions are expected to decay exponentially. c) Paths where the fermions combine into a hydrodynamical operator (local charge $n$ and energy $h$ densities, and their derivatices) are expected to decay as a power law according to hydrodynamics. Contributions of type c) are expected to dominate at late times.}
    \label{fig:twopaths}
\end{figure}

Our numerical simulations suggest that the ballistic-diffusive crossover time and diffusion constant both scale as $\Delta^{-2}$. Here we show that the same scaling is obtained perturbatively using a Fermi's golden rule style of argument, proceeding as follows. We first argue that the lifetime of a single Fermion Green's function goes as $\Delta^{-2}$, and then use that result to deduce that the diffusion constant  has the same scaling. However, this argument is ultimately uncontrolled in ways we specify, showing the need for the numerical methods developed above.

Along the way it is useful to interpret the ballistic-diffusive crossover in an operator-spreading picture \cite{Keyserlingk2022operator}, which we now introduce. The correlation functions used to determine the diffusion constant can all be written as inner products on the space of operators $\langle A(t) | B\rangle$ where $A,B$ are operators, and $A$ is evolved in the Heisenberg picture. This inner product is a quantum transition amplitude in the space of operators (rather than the more familiar space of states). It is the amplitude for an operator $A$ to Heisenberg evolve to an operator $B$ at a time $t$. By Trotterising the Heisenberg time evolution and inserting a complete set of states, we can express $\langle A(t) | B\rangle$ as a sum over operator histories (see Fig.~\ref{fig:G} for a illustration of some such histories). Each history has the same boundary condition: The initial/final configurations are $A,B$ respectively. In the present case, a complete set of states can be represented by a string $o^{\mu}\equiv \otimes_r o_r^{\mu_r} $, where $o^{0,1,2,3}=I,f^\dagger,f,[f^\dagger, f]$ is a complete on-site basis of operators. The fermion creation/annihilation operators $o^{1,2}$ are represented by horizontal lines with orientations $\rightarrow,\leftarrow$ respectively. On the other hand $o^{3}=1-2n$ is represented by a double line (a natural notation given that it is bilinear in fermion operators). $o^{0}$ is represented by the absence of a line (Fig.~\ref{fig:G}). There are rules constraining the allowed histories. For example, unitarity implies the lines cannot start or end in a region of $I$'s. U(1) symmetry implies that difference between the number of $f$'s and $f^\dagger$'s remains constant in time. 

In the absence of interactions, the $f,f^\dagger$ operators evolve into a linear superposition of single $f,f^\dagger$ operators respectively, according to the equation Eq.~\eqref{eq:free_fermion_operator_evolution}. This is a property special to quadratic Hamiltonians. In the corresponding operator spreading picture, this means that one only need sum over histories where a single fermion operator changes position (Fig.~\ref{fig:G}(a)). In the presence of interactions, on the other hand, new operator histories become relevant, where the number of operator worldlines increases or decreases (Fig.~\ref{fig:G}(b)), and the histories become more `complicated' in a sense we describe shortly. 

We now argue that there is a $\Delta^{-2}$ time-scale associated with Fermion operators becoming more complicated. To be more precise, consider the Green's function $G(x,t)\equiv\langle f_{x}(t)|f_{0}(0)\rangle$. This transition amplitude is associated with operator histories that begin with $f$ at position $x$ and end with an $f$ at position $0$ at time $t$. In the absence of interactions $G^{0}(x,t)=K(x,t)$ behaves ballistically; it is supported in a region $|x|<v t$, and has typical si fze $1/\sqrt{t}$, consistent with a sum rule $\sum_x |G^{0}(x,t)|^2=1$. The corresponding simple operator histories are shown in Fig.~\ref{fig:G}(a). However in the presence of interactions, the Green's function is expected to decay much faster (indeed, exponentially) with time~\cite{bohrdt2017scrambling}, and this is associated with the appearance of more complicated diagrams Fig.~\ref{fig:G}(b). The idea (see \cite{Keyserlingk2022operator}) is that once operators are allowed to branch, the initial operator $f_x$ is much more likely to grow in Fermi weight than it is to end up as the Fermi weight $1$ operator $f_{0}$, which implies that the transition amplitude/Green's function decays in time.

We conjecture that in the presence of interactions, the Green's function decays exponentially as $G(x,t)\sim e^{-O(\Delta^2 t)}$. We can argue for this using the \emph{memory matrix} (MM) formalism~\cite{forster2018hydrodynamic}. In the MM formalism, one divides the set of observables into two categories, `fast' and `slow' modes. In our case, we choose the latter to be linear combinations of single-fermion
operators; the fast space is taken to be the orthogonal complement to the slow space. One can then derive an exact equation of motion  for $G$ in momentum space~\cite{forster2018hydrodynamic} which takes the following form:
\begin{equation}
\partial_{t}G_{k}-\mathrm{i} m_{k}G_{k}+\int_{0}^{t}ds \Sigma_{k}(s)G_{k}(t-s)=0.
\end{equation}
Here, $m_{k}$ represents the free (quadratic) part of the dynamics, such that $[H_0,f_k]=-m_k f_k$\footnote{Our Hamiltonian has unit cell of size $2$, so $m,G,\Gamma,\Sigma$ are each $2\times 2$ matrices, and $f_k,f^\dagger_k$ also carry sublattice indices. We suppress the corresponding sublattice index, which is unimportant for our discussion. Note $m_k$ is a Hermitian matrix.}. $\Sigma_k(s)$ on the other hand represents a memory term, arising from operators leaving and then later re-entering the slow subspace. It is defined as
\begin{equation}
\Sigma_{k}(s)=\langle f_{k} | LQe^{-\mathrm{i}QLQs}QL| f_{k}\rangle,
\end{equation}
where $L=[H,\cdot]$ is the many-body Liouvillian generating the dynamics in the Heisenberg picture, and $Q$ is the projector onto the fast-space. Using the fact that the non-interacting terms do not induce any transitions from slow to fast operators, we can further simplify the above expression to
\begin{align*}
\Sigma_{k}(s) =\Delta^{2}\Gamma_{k}(s) \equiv \Delta^{2} \langle [V,f_{k}] |Q e^{-\mathrm{i}QLQ s}Q  | [V,f_{k}]\rangle,
\end{align*}
where $V$ is the interaction term and $\Delta$ is the interaction strength, appearing in the Hamiltonian as $H = H_0 + \Delta V$. It is readily verified that $[V,f_k]$ is a 3-fermion operator, therefore  $\Gamma_{k}(s)$  is a 3-fermion autocorrelation function with dynamics constrained to the fast space (i.e., generated by $QLQ$). Since $\Gamma_{k}$ only involves dynamics in the fast subspace, we expect that it decays quickly in time.  Putting this together, we have
\begin{equation}\label{eq:fg1_pre}
\partial_{t}G_{k}-\mathrm{i}m_{k}G_{k}+\Delta^{2}\int_{0}^{t}ds\Gamma_{k}(s)G_{k}(t-s)=0.
\end{equation}
A Fermi's golden rule scaling emerges from this equation of motion if $\Gamma_{k}$ decays over a time scale $\kappa^{-1}$ that is independent of interactions in the small $\Delta$ limit.  This allows one to truncate the integral in Eq.~\eqref{eq:fg1_pre} to obtain
\begin{equation}
\partial_{t}G_{k}-\mathrm{i}m_{k}G_{k}+\frac{\Delta^{2}}{\kappa}G_{k}\approx 0.
\end{equation}
This implies $\partial_{t}|G_{k}|^{2}\approx-\frac{2\Delta^{2}}{\kappa}\left|G_{k}\right|^{2}$, so that the single-fermion operators have a lifetime scaling
as $\Delta^{-2}$, in agreement with the data presented in Fig.~\ref{fig:crossover}. 

In fact, this argument is delicate, especially in 1D. It turns out that in the non-interacting approximation\footnote{Obtained by replacing $QLQ$ with $QL_0 Q$ in the definition of $\Gamma$ where $L_0=[H_0,\cdot]$.} the
the 3-fermion correlator decays as $|\Gamma_k (s)| \sim 1/s$ (consistent with dimension counting), which is not sufficiently quick to safely truncate the time integral\footnote{Interestingly $|\Gamma_k(s)| \sim 1/s^d$ in spatial dimension $d$, suggesting that truncation is safe in $d>1$.}. A self-consistent resolution to this issue is to include the effects of interactions on $\Gamma$ through the ansatz ${|\Gamma_k (s)| \sim e^{-c \Delta^2 s}/s}$, which has the expected $1/s$ decay at early times, and the expected exponential decay at later times. This ansatz leads to the advertised FGR lifetime for $G_k$ but with a weak additional logarithmic dependence $|G_k(t)|\sim e^{-O(\Delta^2 \log(\Delta^{-1}) t)}$. Our numerical data are insufficiently resolved to confirm or exclude such subtle logarithmic corrections, so henceforth we will ignore them.

The uncontrolled argument above suggests that that fermion operators propagate freely and ballistically up until time $t_{G}\sim 1/\Delta^2$, past which interactions become important. In the operator spreading language, the worldlines of fermion operators become complicated on said timescale. What are the consequences for the diffusion constant? Conflating $t_G$ with a mean free time for particle collisions, and substituting the result into the Drude formula, we expect the diffusion constant to also scale as $t_G$. 

In what follows we make a more explicit argument that $D\sim t_G$. We start from the Kubo formula for the diffusion constant, and evaluate it using an operator spreading picture and the above results. We argue that time derivative of the diffusion constant $\dot{D}(t)$ is $O(1)$ early times $t \lesssim t_G$, but starts to decay algebraically once $t\gtrsim t_G$. As a result $D(t)= D + \mathrm{poly}(t^{-1})$ at late times, where $D$ scales as $t_G$.

The Kubo Formula (equivalent to Eq.~\ref{eq:D_1st} in the thermodynamic limit) implies that the time derivative of $D(t)$, as defined in Eq.~\eqref{eq:D_1st},  is $\dot{D}(t)=\langle \mathcal{J} (t)|\mathcal{J}(0)\rangle/L$, where $\mathcal{J}$ is the total current $\mathcal{J}=\sum_{x}\bigl[\mathrm{i}f_{x}^{\dagger}f_{x+1}+\mathrm{h.c.}\bigr] = \sum_x j_x$, which is quadratic in Fermion operators.  Therefore $\dot{D}(t)$ involves a sum of correlations of the form
\begin{equation} \label{eq:partofJ}
\langle (f^\dagger_{x}f_{x+1})(t)|f^\dagger_{y}f_{y+1}\rangle.
\end{equation}
This expression can be read as a propagator in the space of operators; it represents the amplitudes for the operators $f^{\dagger}_{x} f_{x+1}$ to end up in the configuration $f^\dagger_{y}f_{y+1}$ at time $t$ under Heisenberg dynamics (Fig.~\ref{fig:twopaths}).

At the very earliest times, interactions are unimportant and each
$f,f^{\dagger}$ making up $j_x$ evolves according to Eq.~\eqref{eq:free_fermion_operator_evolution}. 
In this case, it is easy to verify from analysing the propagator $K$ (which corresponds to coherent/ballistic dynamics of the fermion operators)
that $\langle j_{x}(t)j_{y}\rangle\sim1/t$ for $|x-y|<vt$, which
implies $\dot{D}(t)$ is $O(1)$ at early times\footnote{As an aside note that $\mathcal{J}$ is itself a conserved quantity in the non-interacting model when $h=0$, so $\dot{D}$ is a constant in that case.}.  Thus $D(t)$ grows linearly in $t$ at early times. In the operator spreading picture, the fermion operators spread independently and ballistically in this regime, undergoing essentially no branching events (Fig.~\ref{fig:twopaths}(a)). 

However at later times interactions become important, and the fermion operators comprising $\mathcal{J}$ branch into more complicated superpositions of higher Fermi weight operators (Fig.~\ref{fig:twopaths}(b)). As mentioned above, this process is difficult to reverse (in ergodic systems) \cite{Keyserlingk2022operator}. So the amplitude that $f^\dagger_x,f_{x+1}$ separately evolve to $f^\dagger_y,f_{y+1}$ while having many branching events in the interim (Fig.~\ref{fig:twopaths}(b)) ends up decaying exponentially in time as $e^{-O(t/t_G)}$. In other words, the worldines of each fermion operator in the expressions for $\dot{D}(t)$ becomes costly; they develop a linear line tension.

In this later-time regime, a different class of operator history becomes dominant: the paths where the fermion operators making up $j_{x}$ coalesce into hydrodynamical variables (Fig.~\ref{fig:twopaths}(c)); these operators decay more slowly (as a power law in time, rather than exponentially), and so effectively have a lower line tension\footnote{Here it is important that $j$ has a charge-neutral component, otherwise it would not be able to develop an overlap with hydrodynamical slow modes \cite{Keyserlingk2022operator}. This is ultimately the reason why $\langle j_x (t)\mid j_y \rangle$ correlators decay as a power law, while the single-particle Green's function $\langle f_x (t)\mid f_y \rangle$ decays exponentially.}. The simplest and most relevant hydrodynamical operators in ergodic systems involve products and derivatives of the local energy ($h$) and charge ($n$) densities of the form $\partial_x^a n^{a'} \partial^b_x h^{b'}$. The ones relevant to this discussion are $h\partial_{x}n, h^2\partial_{x}n,\ldots$, although which of these are present will generally can depend on discrete symmetries, filling and temperatures \cite{delacretaz_hartnoll19,Mukerjee_2006}. These operators have less costly world-lines, because of the underlying local conservation laws, and their correlations decay as power laws $1/t^{\alpha}$ rather than exponentially\footnote{One always finds $\alpha \geq 3/2$ \cite{delacretaz_hartnoll19,Mukerjee_2006,Keyserlingk2022operator}}. We expect the crossover time (where both types
of path balance) to obey $e^{-t/t_G} \sim C/t^{\alpha}$,
consistent with crossover $t_{D}\sim  t_G$ with possible logarithmic
enhancements.
In conclusion, we expect $\dot{D}(t) = O(1)$ until a time $O(t_G)$ at which point $\dot{D}(t)$ will start to decay as a power law. This suggests that $D\sim t_G$ as expected, and that $D(t)$ approaches this final value as a power law $1/t^{\alpha-1}$.

\section{Conclusions}

In this paper, we investigated the transport of interacting one-dimensional fermions using a novel numerical scheme, which we developed building on the earlier DAOE method~\cite{rakovszky2022dissipation}. This method, which we named fDAOE, evolves fermionic observables in time, applying to them a truncation which discards observables that involve large numbers of creation / annihilation operators, thus making the dynamics tractable. In one dimension, this truncation scheme has a simple representation in a matrix product operator language, which allows for efficient numerical simulations. 

Unlike the existing DAOE method,  fDAOE is exact for free fermions, and we presented numerical evidence that it is able to correctly capture long-time transport properties for a range of interactions strengths. In particular, we focused on a regime where interactions are weak, which results in parametrically long time scales that are difficult to simulate using brute force approaches. We showed that fDAOE can overcome this limitation, and obtain the expected crossover from ballistic to diffusive transport. We used fDAOE extract both the crossover time scale, and the value of the diffusion constant, both scaling as $\Delta^{-2}$ with the interaction strength $\Delta$. We further justified this scaling by outlining a perturbative Fermi's golden rule calculation, suggesting that it is at this $O(\Delta^{-2})$ timescale that the free fermion approximation breaks down. We also provided an operator spreading picture of the crossover to the late-time diffusive regime. 

Our results open the way towards more thorough investigations of charge transport in interacting fermionic systems. There are various directions in which the method could be further extended to bring it closer to describing experimentally relevant systems. One obvious extension would be to two-dimensional systems, where one could for example try to apply fDAOE in the quasi-1D limit of thin cylinders, similarly to other MPS-based algorithms~\cite{stoudenmire2012studying}. Another open problem is to extend the validity of fDAOE from the infinite temperature limit we considered to finite temperatures and study its effect on transport. Finally, one could consider including interactions with phonons~\cite{brockt2015matrix}.

\emph{Note added}. ---  We would like to bring the reader's attention to a related independent work of Kuo, Ware, Lunts, Hafezi, and White (in preparation). 

\begin{acknowledgments}
We thank Romain Vasseur, Sarang Gopalakrishnan, Iliya Essin and Alex Michailidis for useful conversations. CvK is supported by a UKRI Future Leaders Fellowship MR/T040947/1. T.R. is supported in part by the Stanford Q-Farm Bloch Postdoctoral Fellowship in Quantum Science and Engineering. 
\end{acknowledgments}
\appendix

\section{Convergence with bond dimension and $\ell_*$}\label{app:convergence}

\begin{figure*}
    \centering
    \includegraphics[width=\textwidth]{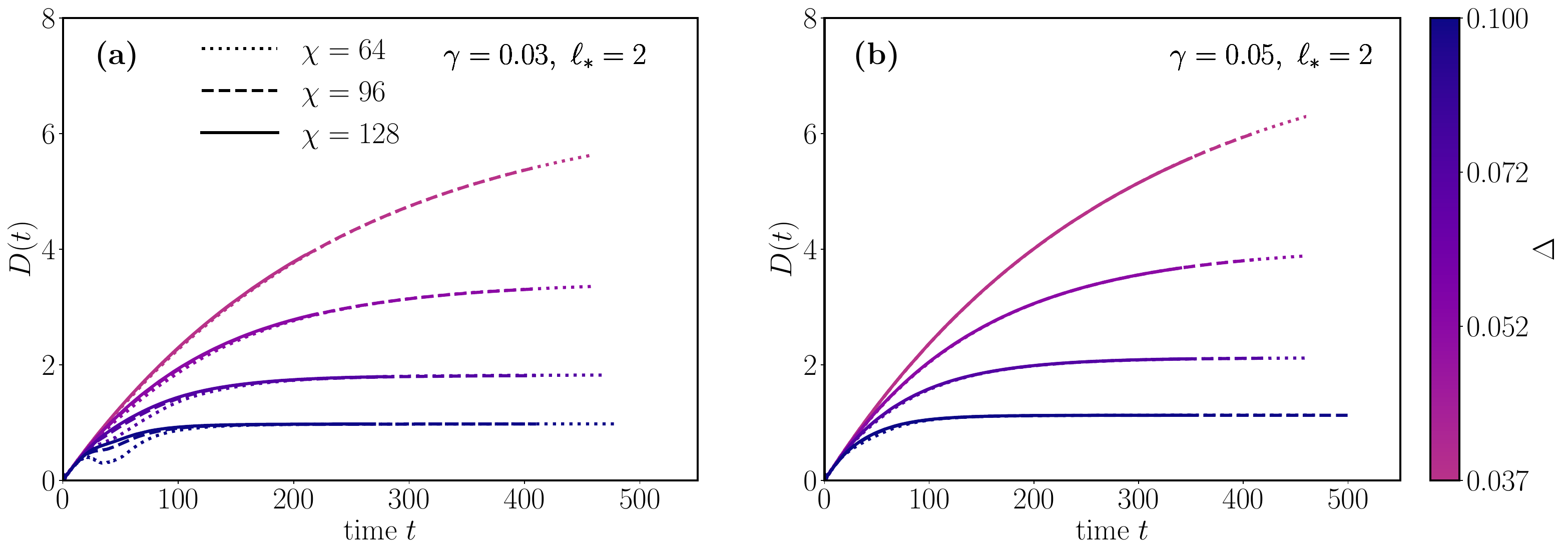}
    \caption{\textbf{(a)} For $\ell_*=2$, we plot $D(t)$ for different bond dimensions, $\chi=64$ (dotted line), $\chi=96$ (dashed), $\chi=128$ (solid), and $\gamma=0.03$. For larger $\Delta$, the data is less well converged at early times, but converged at late times in the diffusive regime. \textbf{(b)} Same plot for $\gamma=0.05$.}
    \label{fig:conv2}
\end{figure*}

Here we report additional data for different $\ell_*$ and different bond dimensions $\chi = 64, 96, 128$ ($\ell_* = 2$) and $\chi = 256, 512$ ($\ell_*=4$). In Fig.~\ref{fig:conv2} we plot the time-dependent diffusion constant $D(t)$ for $\ell_*=2$ and different bond dimensions, for $\gamma=0.03$ in \ref{fig:conv2}(a) $\gamma=0.05$ in \ref{fig:conv2}(b). We show show data for the four largest $\Delta$ considered in the main text: for decreasing $\Delta$, or increasing $\gamma$, entanglement growth is reduced and errors are smaller. We see that the most notable errors are for $\chi=64$, $\Delta = 0.1$ and $\gamma=0.03$ at early times, as expected. However, even in this case, the late time diffusive dynamics is well converged with the larger bond dimension $\chi=128$. The diffusive transport appears to `recover' from early-time errors as DAOE continuously suppresses entanglement growth. We conclude that the data used in the main text ($\chi=96$) is reasonably well converged in bond dimension, and the extrapolated diffusion constants in Fig.~\ref{fig:diffusion} are not affected by truncation errors.

\begin{figure*}
    \centering
    \includegraphics[width=\textwidth]{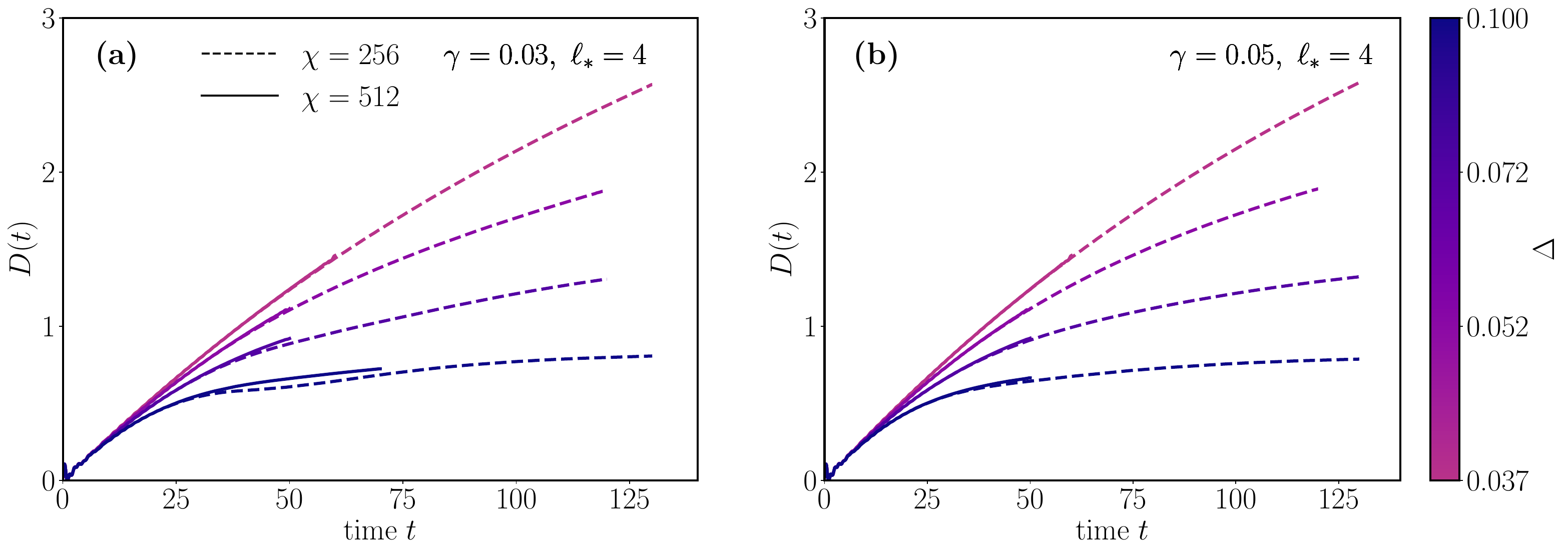}
    \caption{\textbf{(a)} For $\ell_*=4$, same as Fig.~\ref{fig:conv2} with $\chi=256$ (dashed line), $\chi=512$ (solid), and $\gamma=0.03$. The convergence is slightly worse compared to the $\ell_*=2$ case. \textbf{(b)} Same plot for $\gamma=0.05$.}
    \label{fig:conv4} 
\end{figure*}

We show the same data but for $\ell_*=4$ in Fig.~\ref{fig:conv4}. Here the story is similar although the truncation errors are larger, as expected from the fact that interactions are less strongly suppressed. Given the large bond dimensions (256 and 512), we are limited to shorter times and it is difficult to assess the convergence for late times. Note that we do not use the $\ell_*=4$ data anywhere in the main text however. 

\begin{figure}
    \centering
    \includegraphics[width=\columnwidth]{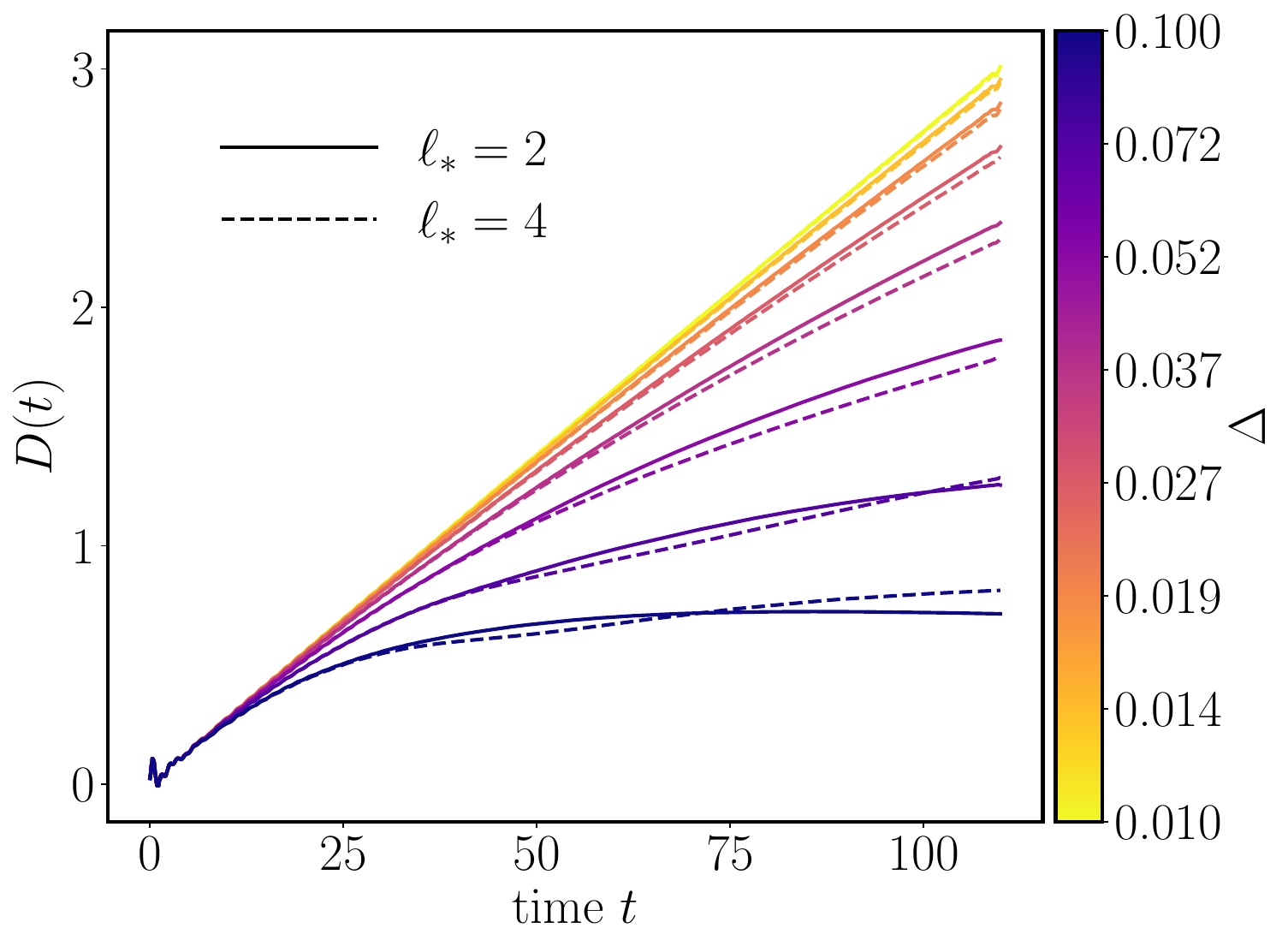}
    \caption{$D(t)$ for $\ell_*=2, \chi=128$ (solid lines) vs.~$\ell_*=4, \chi=256$ (dashed lines), for the different $\Delta$ used in the main text. The $D(t)$ curves are extrapolated in $\gamma \in (0.03, 0.05, 0.1, 0.2)$ for $\Delta\neq0.1$ and $\gamma \in (0.05, 0.1, 0.2)$ for $\Delta=0.1$, due to the larger truncation errors for $\gamma=0.03$ and $\Delta=0.1$.}
    \label{fig:convl}
\end{figure}

To check the convergence between $\ell_*=2$ and $\ell_*=4$, we plot $D(t)$ curves extrapolated in $\gamma\to 0$ with a quadratic fit in Fig.~\ref{fig:convl}, for $\chi=128$ ($\ell_*=2$) and $\chi=256$ ($\ell_*=4$). For $\Delta \neq 0.1$ we extrapolate in $\gamma \in (0.03, 0.05, 0.1, 0.2)$, while for $\Delta = 0.1$ we ignore the 0.03 data point, due to the visible truncation effects in Figs.~\ref{fig:conv2},\ref{fig:conv4}. We obtain a reasonable quantitative convergence, i.e.~ the difference between $\ell_*=2,4$ is much smaller than the difference between different $\Delta$. The curves are not necessarily expected to agree for finite $\gamma$ and we attribute the main error in convergence to the small $\gamma$ set we perform the extrapolation on. 


\section{Entanglement dynamics and diffusive Zeno effect}
\label{app:zeno}
\begin{figure}
    \centering
    \includegraphics[width=\columnwidth]{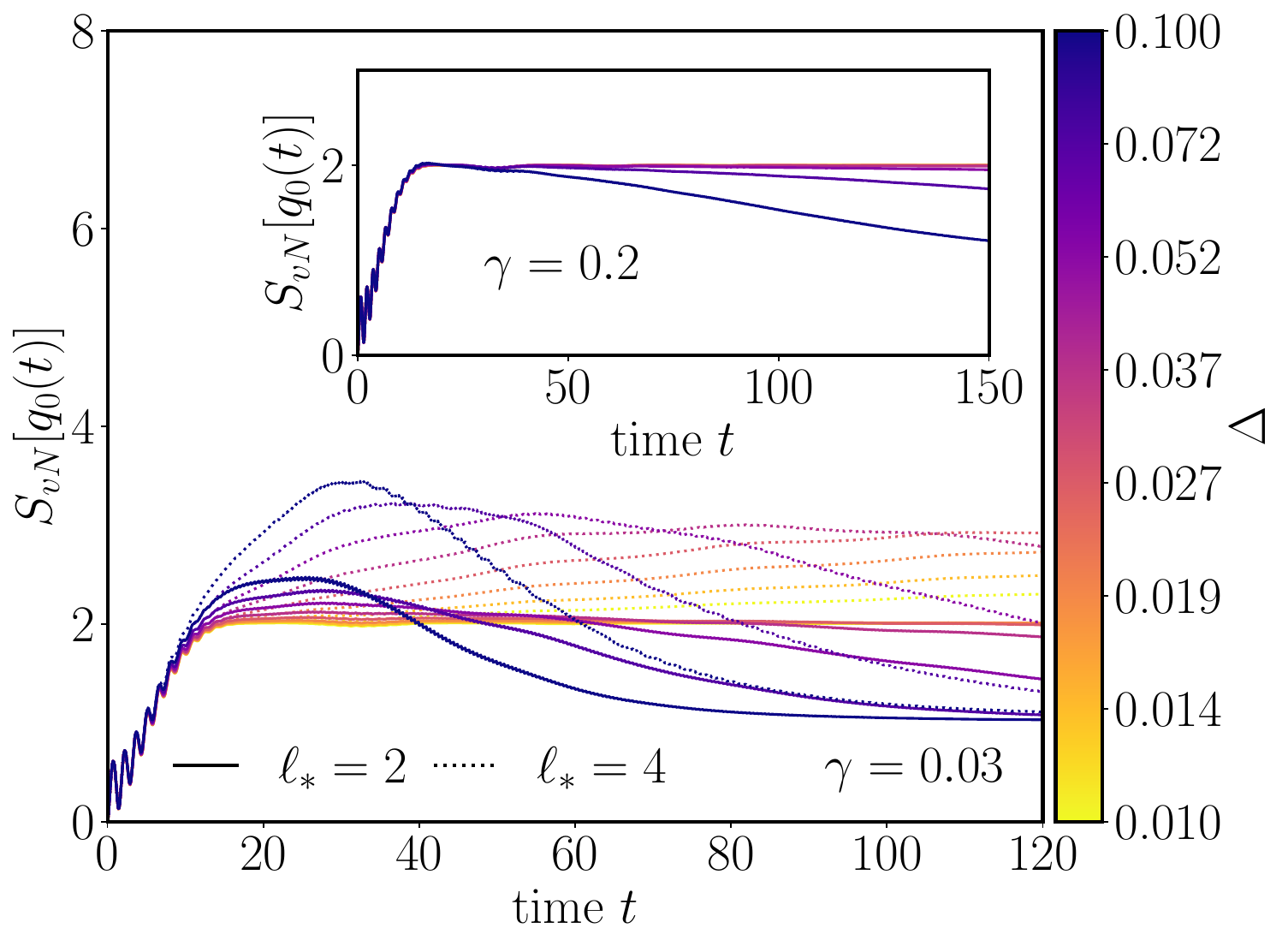}
    \caption{Half chain operator entanglement growth for $\gamma = 0.03$ and $\ell_*=2$ (solid lines), $\ell_*=4$ (dotted lines). Entanglement rapidly approaches a $2\log2$ plateau, before slowly increasing at a rate set by $\Delta$ (for $\ell_*=4$) and eventually decreasing due to the DAOE dissipation. \textbf{Inset:} entanglement growth for $\gamma=0.2$. The stronger dissipation leads to frozen dynamics, with entanglement decaying more slowly compared to the $\gamma=0.03$ case.}
    \label{fig:zeno}
\end{figure}

In this appendix we study the dynamics of operator space entanglement entropy (OSSE) growth under fDAOE, and give an argument for the `diffusive Zeno effect' mentioned in the main text.

We define the operator entanglement of a Heisenberg-evolved operator $q(t)$ as the half-chain von Neumann entropy of (normalised) MPS representation of the operator, $\ket{q(t)}$. The half-chain OSSE essentially measures how correlated the left and right halves of the system are. Under generic unitary evolution, the entanglement grows linearly in time, and storing this highly-entangled state is the main obstacle to `exact' evolution schemes like TEBD. 

Applying the DAOE dissipator decreases this entanglement, and its late time value remains bounded by a cutoff depending on the parameters $\ell_*$ and $\gamma$. This suppression is the reason for the success of DAOE in simulating long-time transport [REF]. 

If Fig.~\ref{fig:zeno} we plot the half-chain OSSE (normalized by $\log2$) for the time-evolved operators $q_x(t) = \sigma_x^z(t)$ we study in the main text, for different $\Delta$. The solid lines are with $\ell_*=2$ and the dotted lines with $\ell_*=4$, and $\gamma=0.03$ in the main figure. We observe a rapid linear growth towards the value $S_{vN} = 2\log 2$, after which the entanglement dynamics depends on $\ell_*$ and the interaction strength $\Delta$. The $2\log2$ value can be understood as follows. $\sigma_x^z$ is bilinear in fermion operators. At early times, the two fermion operators essentially evolve under free dynamics, each becoming a superposition of single fermions within a light cone. Each of these fermion operators has entropy $\sim \log2$ across the central cut, while the entropy of their product $\sigma_x^z(t)$ is the sum of their entropies, namely $2\log2$. 

For $\ell_*=2$, the entanglement plateaus around the free-fermion value $2\log2$ up to a timescale that decreases with $\Delta$, before slowly decaying to $S_{vN} = \log 2$ consistent with diffusion \cite{rakovszky2022dissipation}. This decay is a result of DAOE truncating on the operator space. For $\ell_*=4$, operators are allowed to grow into larger strings (of Fermi weight $\ell=4$) before truncation begins. This leads to a slow growth of entanglement away from the $2\log2$ plateau, with faster rate of entanglement production for larger $\Delta$: at late times, DAOE overcomes the rate of entanglement generation and $S_{vN}$ falls. The slow entanglement growth for weakly interacting systems was the reason to expect that TEBD may be efficient for simulating diffusive transport as $\Delta \to 0$. However, as we showed in Fig.~\ref{fig:crossover}(b), the divergence of the crossover timescale $t_D \sim \Delta^{-2}$ ultimately prevents approaching this regime before entanglement growth becomes an issue. 

We now discuss why the diffusion constant is observed to increase with $\gamma$, as we show in the inset to Fig.~\ref{fig:diffusion}. Again, a naive picture of DAOE as `dissipation' would suggest that the method destroys the system coherence and stabilises diffusion. In the inset to Fig.~\ref{app:zeno} we plot the OSEE growth for $\ell_*=2$, but now for larger $\gamma=0.2$. The entanglement is essentially `frozen' at the free-fermion entanglement value $S_{vN}=2\log 2$, with the entanglement in the case of strong $\Delta$  decaying much slower that for the smaller $\gamma$ case. When the dissipation is strong, the operator is prevented from ever leaving the free-fermion subspace of length $\ell=2$ operators. This can be interpreted as a freezing of the interactions, $\Delta \to 0$, which leads to an diverging diffusion constant and purely ballistic dynamics in the limit $\gamma \to \infty$. We dub this phenomena the `diffusive Zeno effect' as it is suggestive of the phenomena where a repeatedly measured state exhibits frozen evolution.

\bibliography{bib}

\begin{thebibliography}{36}%
\makeatletter
\providecommand \@ifxundefined [1]{%
 \@ifx{#1\undefined}
}%
\providecommand \@ifnum [1]{%
 \ifnum #1\expandafter \@firstoftwo
 \else \expandafter \@secondoftwo
 \fi
}%
\providecommand \@ifx [1]{%
 \ifx #1\expandafter \@firstoftwo
 \else \expandafter \@secondoftwo
 \fi
}%
\providecommand \natexlab [1]{#1}%
\providecommand \enquote  [1]{``#1''}%
\providecommand \bibnamefont  [1]{#1}%
\providecommand \bibfnamefont [1]{#1}%
\providecommand \citenamefont [1]{#1}%
\providecommand \href@noop [0]{\@secondoftwo}%
\providecommand \href [0]{\begingroup \@sanitize@url \@href}%
\providecommand \@href[1]{\@@startlink{#1}\@@href}%
\providecommand \@@href[1]{\endgroup#1\@@endlink}%
\providecommand \@sanitize@url [0]{\catcode `\\12\catcode `\$12\catcode `\&12\catcode `\#12\catcode `\^12\catcode `\_12\catcode `\%12\relax}%
\providecommand \@@startlink[1]{}%
\providecommand \@@endlink[0]{}%
\providecommand \url  [0]{\begingroup\@sanitize@url \@url }%
\providecommand \@url [1]{\endgroup\@href {#1}{\urlprefix }}%
\providecommand \urlprefix  [0]{URL }%
\providecommand \Eprint [0]{\href }%
\providecommand \doibase [0]{https://doi.org/}%
\providecommand \selectlanguage [0]{\@gobble}%
\providecommand \bibinfo  [0]{\@secondoftwo}%
\providecommand \bibfield  [0]{\@secondoftwo}%
\providecommand \translation [1]{[#1]}%
\providecommand \BibitemOpen [0]{}%
\providecommand \bibitemStop [0]{}%
\providecommand \bibitemNoStop [0]{.\EOS\space}%
\providecommand \EOS [0]{\spacefactor3000\relax}%
\providecommand \BibitemShut  [1]{\csname bibitem#1\endcsname}%
\let\auto@bib@innerbib\@empty
\bibitem [{\citenamefont {D'Alessio}\ \emph {et~al.}(2016)\citenamefont {D'Alessio}, \citenamefont {Kafri}, \citenamefont {Polkovnikov},\ and\ \citenamefont {Rigol}}]{RigolReview}%
  \BibitemOpen
  \bibfield  {author} {\bibinfo {author} {\bibfnamefont {L.}~\bibnamefont {D'Alessio}}, \bibinfo {author} {\bibfnamefont {Y.}~\bibnamefont {Kafri}}, \bibinfo {author} {\bibfnamefont {A.}~\bibnamefont {Polkovnikov}},\ and\ \bibinfo {author} {\bibfnamefont {M.}~\bibnamefont {Rigol}},\ }\bibfield  {title} {\bibinfo {title} {From quantum chaos and eigenstate thermalization to statistical mechanics and thermodynamics},\ }\href@noop {} {\bibfield  {journal} {\bibinfo  {journal} {Advances in Physics}\ }\textbf {\bibinfo {volume} {65}},\ \bibinfo {pages} {239} (\bibinfo {year} {2016})}\BibitemShut {NoStop}%
\bibitem [{\citenamefont {Rakovszky}\ \emph {et~al.}(2018)\citenamefont {Rakovszky}, \citenamefont {Pollmann},\ and\ \citenamefont {Von~Keyserlingk}}]{rakovszky2018diffusive}%
  \BibitemOpen
  \bibfield  {author} {\bibinfo {author} {\bibfnamefont {T.}~\bibnamefont {Rakovszky}}, \bibinfo {author} {\bibfnamefont {F.}~\bibnamefont {Pollmann}},\ and\ \bibinfo {author} {\bibfnamefont {C.}~\bibnamefont {Von~Keyserlingk}},\ }\bibfield  {title} {\bibinfo {title} {Diffusive hydrodynamics of out-of-time-ordered correlators with charge conservation},\ }\href@noop {} {\bibfield  {journal} {\bibinfo  {journal} {Physical Review X}\ }\textbf {\bibinfo {volume} {8}},\ \bibinfo {pages} {031058} (\bibinfo {year} {2018})}\BibitemShut {NoStop}%
\bibitem [{\citenamefont {Khemani}\ \emph {et~al.}(2018)\citenamefont {Khemani}, \citenamefont {Vishwanath},\ and\ \citenamefont {Huse}}]{khemani2018operator}%
  \BibitemOpen
  \bibfield  {author} {\bibinfo {author} {\bibfnamefont {V.}~\bibnamefont {Khemani}}, \bibinfo {author} {\bibfnamefont {A.}~\bibnamefont {Vishwanath}},\ and\ \bibinfo {author} {\bibfnamefont {D.~A.}\ \bibnamefont {Huse}},\ }\bibfield  {title} {\bibinfo {title} {Operator spreading and the emergence of dissipative hydrodynamics under unitary evolution with conservation laws},\ }\href@noop {} {\bibfield  {journal} {\bibinfo  {journal} {Physical Review X}\ }\textbf {\bibinfo {volume} {8}},\ \bibinfo {pages} {031057} (\bibinfo {year} {2018})}\BibitemShut {NoStop}%
\bibitem [{\citenamefont {Parker}\ \emph {et~al.}(2019)\citenamefont {Parker}, \citenamefont {Cao}, \citenamefont {Avdoshkin}, \citenamefont {Scaffidi},\ and\ \citenamefont {Altman}}]{parker2019universal}%
  \BibitemOpen
  \bibfield  {author} {\bibinfo {author} {\bibfnamefont {D.~E.}\ \bibnamefont {Parker}}, \bibinfo {author} {\bibfnamefont {X.}~\bibnamefont {Cao}}, \bibinfo {author} {\bibfnamefont {A.}~\bibnamefont {Avdoshkin}}, \bibinfo {author} {\bibfnamefont {T.}~\bibnamefont {Scaffidi}},\ and\ \bibinfo {author} {\bibfnamefont {E.}~\bibnamefont {Altman}},\ }\bibfield  {title} {\bibinfo {title} {A universal operator growth hypothesis},\ }\href@noop {} {\bibfield  {journal} {\bibinfo  {journal} {Physical Review X}\ }\textbf {\bibinfo {volume} {9}},\ \bibinfo {pages} {041017} (\bibinfo {year} {2019})}\BibitemShut {NoStop}%
\bibitem [{\citenamefont {Lucas}(2019)}]{lucas2019operator}%
  \BibitemOpen
  \bibfield  {author} {\bibinfo {author} {\bibfnamefont {A.}~\bibnamefont {Lucas}},\ }\bibfield  {title} {\bibinfo {title} {Operator size at finite temperature and planckian bounds on quantum dynamics},\ }\href@noop {} {\bibfield  {journal} {\bibinfo  {journal} {Physical review letters}\ }\textbf {\bibinfo {volume} {122}},\ \bibinfo {pages} {216601} (\bibinfo {year} {2019})}\BibitemShut {NoStop}%
\bibitem [{\citenamefont {Forster}(2018)}]{forster2018hydrodynamic}%
  \BibitemOpen
  \bibfield  {author} {\bibinfo {author} {\bibfnamefont {D.}~\bibnamefont {Forster}},\ }\href@noop {} {\emph {\bibinfo {title} {Hydrodynamic fluctuations, broken symmetry, and correlation functions}}}\ (\bibinfo  {publisher} {CRC Press},\ \bibinfo {year} {2018})\BibitemShut {NoStop}%
\bibitem [{\citenamefont {Rakovszky}\ \emph {et~al.}(2022)\citenamefont {Rakovszky}, \citenamefont {von Keyserlingk},\ and\ \citenamefont {Pollmann}}]{rakovszky2022dissipation}%
  \BibitemOpen
  \bibfield  {author} {\bibinfo {author} {\bibfnamefont {T.}~\bibnamefont {Rakovszky}}, \bibinfo {author} {\bibfnamefont {C.}~\bibnamefont {von Keyserlingk}},\ and\ \bibinfo {author} {\bibfnamefont {F.}~\bibnamefont {Pollmann}},\ }\bibfield  {title} {\bibinfo {title} {Dissipation-assisted operator evolution method for capturing hydrodynamic transport},\ }\href@noop {} {\bibfield  {journal} {\bibinfo  {journal} {Physical Review B}\ }\textbf {\bibinfo {volume} {105}},\ \bibinfo {pages} {075131} (\bibinfo {year} {2022})}\BibitemShut {NoStop}%
\bibitem [{\citenamefont {Klein~Kvorning}\ \emph {et~al.}(2022)\citenamefont {Klein~Kvorning}, \citenamefont {Herviou},\ and\ \citenamefont {Bardarson}}]{klein2022time}%
  \BibitemOpen
  \bibfield  {author} {\bibinfo {author} {\bibfnamefont {T.}~\bibnamefont {Klein~Kvorning}}, \bibinfo {author} {\bibfnamefont {L.}~\bibnamefont {Herviou}},\ and\ \bibinfo {author} {\bibfnamefont {J.~H.}\ \bibnamefont {Bardarson}},\ }\bibfield  {title} {\bibinfo {title} {Time-evolution of local information: thermalization dynamics of local observables},\ }\href@noop {} {\bibfield  {journal} {\bibinfo  {journal} {SciPost Physics}\ }\textbf {\bibinfo {volume} {13}},\ \bibinfo {pages} {080} (\bibinfo {year} {2022})}\BibitemShut {NoStop}%
\bibitem [{\citenamefont {Artiaco}\ \emph {et~al.}(2023)\citenamefont {Artiaco}, \citenamefont {Fleckenstein}, \citenamefont {Aceituno}, \citenamefont {Kvorning},\ and\ \citenamefont {Bardarson}}]{artiaco2023efficient}%
  \BibitemOpen
  \bibfield  {author} {\bibinfo {author} {\bibfnamefont {C.}~\bibnamefont {Artiaco}}, \bibinfo {author} {\bibfnamefont {C.}~\bibnamefont {Fleckenstein}}, \bibinfo {author} {\bibfnamefont {D.}~\bibnamefont {Aceituno}}, \bibinfo {author} {\bibfnamefont {T.~K.}\ \bibnamefont {Kvorning}},\ and\ \bibinfo {author} {\bibfnamefont {J.~H.}\ \bibnamefont {Bardarson}},\ }\bibfield  {title} {\bibinfo {title} {Efficient large-scale many-body quantum dynamics via local-information time evolution},\ }\href@noop {} {\bibfield  {journal} {\bibinfo  {journal} {arXiv preprint arXiv:2310.06036}\ } (\bibinfo {year} {2023})}\BibitemShut {NoStop}%
\bibitem [{\citenamefont {White}(2023)}]{white2023effective}%
  \BibitemOpen
  \bibfield  {author} {\bibinfo {author} {\bibfnamefont {C.~D.}\ \bibnamefont {White}},\ }\bibfield  {title} {\bibinfo {title} {Effective dissipation rate in a liouvillian-graph picture of high-temperature quantum hydrodynamics},\ }\href@noop {} {\bibfield  {journal} {\bibinfo  {journal} {Physical Review B}\ }\textbf {\bibinfo {volume} {107}},\ \bibinfo {pages} {094311} (\bibinfo {year} {2023})}\BibitemShut {NoStop}%
\bibitem [{\citenamefont {Von~Keyserlingk}\ \emph {et~al.}(2022)\citenamefont {Von~Keyserlingk}, \citenamefont {Pollmann},\ and\ \citenamefont {Rakovszky}}]{Keyserlingk2022operator}%
  \BibitemOpen
  \bibfield  {author} {\bibinfo {author} {\bibfnamefont {C.}~\bibnamefont {Von~Keyserlingk}}, \bibinfo {author} {\bibfnamefont {F.}~\bibnamefont {Pollmann}},\ and\ \bibinfo {author} {\bibfnamefont {T.}~\bibnamefont {Rakovszky}},\ }\bibfield  {title} {\bibinfo {title} {Operator backflow and the classical simulation of quantum transport},\ }\href@noop {} {\bibfield  {journal} {\bibinfo  {journal} {Physical Review B}\ }\textbf {\bibinfo {volume} {105}},\ \bibinfo {pages} {245101} (\bibinfo {year} {2022})}\BibitemShut {NoStop}%
\bibitem [{\citenamefont {Hartnoll}\ and\ \citenamefont {Mackenzie}(2022)}]{strange_metals_sean}%
  \BibitemOpen
  \bibfield  {author} {\bibinfo {author} {\bibfnamefont {S.~A.}\ \bibnamefont {Hartnoll}}\ and\ \bibinfo {author} {\bibfnamefont {A.~P.}\ \bibnamefont {Mackenzie}},\ }\bibfield  {title} {\bibinfo {title} {Colloquium: Planckian dissipation in metals},\ }\href {https://doi.org/10.1103/RevModPhys.94.041002} {\bibfield  {journal} {\bibinfo  {journal} {Rev. Mod. Phys.}\ }\textbf {\bibinfo {volume} {94}},\ \bibinfo {pages} {041002} (\bibinfo {year} {2022})}\BibitemShut {NoStop}%
\bibitem [{\citenamefont {Ott}\ \emph {et~al.}(2004)\citenamefont {Ott}, \citenamefont {de~Mirandes}, \citenamefont {Ferlaino}, \citenamefont {Roati}, \citenamefont {Modugno},\ and\ \citenamefont {Inguscio}}]{noneq_fermi_hubbard_ott}%
  \BibitemOpen
  \bibfield  {author} {\bibinfo {author} {\bibfnamefont {H.}~\bibnamefont {Ott}}, \bibinfo {author} {\bibfnamefont {E.}~\bibnamefont {de~Mirandes}}, \bibinfo {author} {\bibfnamefont {F.}~\bibnamefont {Ferlaino}}, \bibinfo {author} {\bibfnamefont {G.}~\bibnamefont {Roati}}, \bibinfo {author} {\bibfnamefont {G.}~\bibnamefont {Modugno}},\ and\ \bibinfo {author} {\bibfnamefont {M.}~\bibnamefont {Inguscio}},\ }\bibfield  {title} {\bibinfo {title} {Collisionally induced transport in periodic potentials},\ }\href {https://doi.org/10.1103/PhysRevLett.92.160601} {\bibfield  {journal} {\bibinfo  {journal} {Phys. Rev. Lett.}\ }\textbf {\bibinfo {volume} {92}},\ \bibinfo {pages} {160601} (\bibinfo {year} {2004})}\BibitemShut {NoStop}%
\bibitem [{\citenamefont {Scherg}\ \emph {et~al.}(2018)\citenamefont {Scherg}, \citenamefont {Kohlert}, \citenamefont {Herbrych}, \citenamefont {Stolpp}, \citenamefont {Bordia}, \citenamefont {Schneider}, \citenamefont {Heidrich-Meisner}, \citenamefont {Bloch},\ and\ \citenamefont {Aidelsburger}}]{noneq_fermi_hubbard_monika}%
  \BibitemOpen
  \bibfield  {author} {\bibinfo {author} {\bibfnamefont {S.}~\bibnamefont {Scherg}}, \bibinfo {author} {\bibfnamefont {T.}~\bibnamefont {Kohlert}}, \bibinfo {author} {\bibfnamefont {J.}~\bibnamefont {Herbrych}}, \bibinfo {author} {\bibfnamefont {J.}~\bibnamefont {Stolpp}}, \bibinfo {author} {\bibfnamefont {P.}~\bibnamefont {Bordia}}, \bibinfo {author} {\bibfnamefont {U.}~\bibnamefont {Schneider}}, \bibinfo {author} {\bibfnamefont {F.}~\bibnamefont {Heidrich-Meisner}}, \bibinfo {author} {\bibfnamefont {I.}~\bibnamefont {Bloch}},\ and\ \bibinfo {author} {\bibfnamefont {M.}~\bibnamefont {Aidelsburger}},\ }\bibfield  {title} {\bibinfo {title} {Nonequilibrium mass transport in the 1d fermi-hubbard model},\ }\href {https://doi.org/10.1103/PhysRevLett.121.130402} {\bibfield  {journal} {\bibinfo  {journal} {Phys. Rev. Lett.}\ }\textbf {\bibinfo {volume} {121}},\ \bibinfo {pages} {130402} (\bibinfo {year} {2018})}\BibitemShut {NoStop}%
\bibitem [{\citenamefont {Nichols}\ \emph {et~al.}(2019)\citenamefont {Nichols}, \citenamefont {Cheuk}, \citenamefont {Okan}, \citenamefont {Hartke}, \citenamefont {Mendez}, \citenamefont {Senthil}, \citenamefont {Khatami}, \citenamefont {Zhang},\ and\ \citenamefont {Zwierlein}}]{strange_metal_cold_atom1}%
  \BibitemOpen
  \bibfield  {author} {\bibinfo {author} {\bibfnamefont {M.~A.}\ \bibnamefont {Nichols}}, \bibinfo {author} {\bibfnamefont {L.~W.}\ \bibnamefont {Cheuk}}, \bibinfo {author} {\bibfnamefont {M.}~\bibnamefont {Okan}}, \bibinfo {author} {\bibfnamefont {T.~R.}\ \bibnamefont {Hartke}}, \bibinfo {author} {\bibfnamefont {E.}~\bibnamefont {Mendez}}, \bibinfo {author} {\bibfnamefont {T.}~\bibnamefont {Senthil}}, \bibinfo {author} {\bibfnamefont {E.}~\bibnamefont {Khatami}}, \bibinfo {author} {\bibfnamefont {H.}~\bibnamefont {Zhang}},\ and\ \bibinfo {author} {\bibfnamefont {M.~W.}\ \bibnamefont {Zwierlein}},\ }\bibfield  {title} {\bibinfo {title} {Spin transport in a mott insulator of ultracold fermions},\ }\href {https://doi.org/10.1126/science.aat4387} {\bibfield  {journal} {\bibinfo  {journal} {Science}\ }\textbf {\bibinfo {volume} {363}},\ \bibinfo {pages} {383} (\bibinfo {year} {2019})},\ \Eprint {https://arxiv.org/abs/https://www.science.org/doi/pdf/10.1126/science.aat4387}
  {https://www.science.org/doi/pdf/10.1126/science.aat4387} \BibitemShut {NoStop}%
\bibitem [{\citenamefont {Brown}\ \emph {et~al.}(2019)\citenamefont {Brown}, \citenamefont {Mitra}, \citenamefont {Guardado-Sanchez}, \citenamefont {Nourafkan}, \citenamefont {Reymbaut}, \citenamefont {Hébert}, \citenamefont {Bergeron}, \citenamefont {Tremblay}, \citenamefont {Kokalj}, \citenamefont {Huse}, \citenamefont {Schauß},\ and\ \citenamefont {Bakr}}]{strange_metal_cold_atom2}%
  \BibitemOpen
  \bibfield  {author} {\bibinfo {author} {\bibfnamefont {P.~T.}\ \bibnamefont {Brown}}, \bibinfo {author} {\bibfnamefont {D.}~\bibnamefont {Mitra}}, \bibinfo {author} {\bibfnamefont {E.}~\bibnamefont {Guardado-Sanchez}}, \bibinfo {author} {\bibfnamefont {R.}~\bibnamefont {Nourafkan}}, \bibinfo {author} {\bibfnamefont {A.}~\bibnamefont {Reymbaut}}, \bibinfo {author} {\bibfnamefont {C.-D.}\ \bibnamefont {Hébert}}, \bibinfo {author} {\bibfnamefont {S.}~\bibnamefont {Bergeron}}, \bibinfo {author} {\bibfnamefont {A.-M.~S.}\ \bibnamefont {Tremblay}}, \bibinfo {author} {\bibfnamefont {J.}~\bibnamefont {Kokalj}}, \bibinfo {author} {\bibfnamefont {D.~A.}\ \bibnamefont {Huse}}, \bibinfo {author} {\bibfnamefont {P.}~\bibnamefont {Schauß}},\ and\ \bibinfo {author} {\bibfnamefont {W.~S.}\ \bibnamefont {Bakr}},\ }\bibfield  {title} {\bibinfo {title} {Bad metallic transport in a cold atom fermi-hubbard system},\ }\href {https://doi.org/10.1126/science.aat4134} {\bibfield  {journal} {\bibinfo  {journal} {Science}\ }\textbf
  {\bibinfo {volume} {363}},\ \bibinfo {pages} {379} (\bibinfo {year} {2019})},\ \Eprint {https://arxiv.org/abs/https://www.science.org/doi/pdf/10.1126/science.aat4134} {https://www.science.org/doi/pdf/10.1126/science.aat4134} \BibitemShut {NoStop}%
\bibitem [{\citenamefont {Kardar}(2007)}]{kardar_book_particles}%
  \BibitemOpen
  \bibfield  {author} {\bibinfo {author} {\bibfnamefont {M.}~\bibnamefont {Kardar}},\ }\href@noop {} {\emph {\bibinfo {title} {Statistical physics of particles}}}\ (\bibinfo  {publisher} {Cambridge University Press},\ \bibinfo {address} {Cambridge},\ \bibinfo {year} {2007})\BibitemShut {NoStop}%
\bibitem [{\citenamefont {Ashcroft}\ and\ \citenamefont {Mermin}(2022)}]{ashcroft2022solid}%
  \BibitemOpen
  \bibfield  {author} {\bibinfo {author} {\bibfnamefont {N.~W.}\ \bibnamefont {Ashcroft}}\ and\ \bibinfo {author} {\bibfnamefont {N.~D.}\ \bibnamefont {Mermin}},\ }\href@noop {} {\emph {\bibinfo {title} {Solid state physics}}}\ (\bibinfo  {publisher} {Cengage Learning},\ \bibinfo {year} {2022})\BibitemShut {NoStop}%
\bibitem [{\citenamefont {Schollw{\"o}ck}(2011)}]{schollwock2011density}%
  \BibitemOpen
  \bibfield  {author} {\bibinfo {author} {\bibfnamefont {U.}~\bibnamefont {Schollw{\"o}ck}},\ }\bibfield  {title} {\bibinfo {title} {The density-matrix renormalization group in the age of matrix product states},\ }\href@noop {} {\bibfield  {journal} {\bibinfo  {journal} {Annals of physics}\ }\textbf {\bibinfo {volume} {326}},\ \bibinfo {pages} {96} (\bibinfo {year} {2011})}\BibitemShut {NoStop}%
\bibitem [{\citenamefont {Paeckel}\ \emph {et~al.}(2019)\citenamefont {Paeckel}, \citenamefont {K{\"o}hler}, \citenamefont {Swoboda}, \citenamefont {Manmana}, \citenamefont {Schollw{\"o}ck},\ and\ \citenamefont {Hubig}}]{paeckel2019time}%
  \BibitemOpen
  \bibfield  {author} {\bibinfo {author} {\bibfnamefont {S.}~\bibnamefont {Paeckel}}, \bibinfo {author} {\bibfnamefont {T.}~\bibnamefont {K{\"o}hler}}, \bibinfo {author} {\bibfnamefont {A.}~\bibnamefont {Swoboda}}, \bibinfo {author} {\bibfnamefont {S.~R.}\ \bibnamefont {Manmana}}, \bibinfo {author} {\bibfnamefont {U.}~\bibnamefont {Schollw{\"o}ck}},\ and\ \bibinfo {author} {\bibfnamefont {C.}~\bibnamefont {Hubig}},\ }\bibfield  {title} {\bibinfo {title} {Time-evolution methods for matrix-product states},\ }\href@noop {} {\bibfield  {journal} {\bibinfo  {journal} {Annals of Physics}\ }\textbf {\bibinfo {volume} {411}},\ \bibinfo {pages} {167998} (\bibinfo {year} {2019})}\BibitemShut {NoStop}%
\bibitem [{\citenamefont {Prosen}\ and\ \citenamefont {{\v{Z}}nidari{\v{c}}}(2007)}]{prosen2007efficiency}%
  \BibitemOpen
  \bibfield  {author} {\bibinfo {author} {\bibfnamefont {T.}~\bibnamefont {Prosen}}\ and\ \bibinfo {author} {\bibfnamefont {M.}~\bibnamefont {{\v{Z}}nidari{\v{c}}}},\ }\bibfield  {title} {\bibinfo {title} {Is the efficiency of classical simulations of quantum dynamics related to integrability?},\ }\href@noop {} {\bibfield  {journal} {\bibinfo  {journal} {Physical Review E}\ }\textbf {\bibinfo {volume} {75}},\ \bibinfo {pages} {015202} (\bibinfo {year} {2007})}\BibitemShut {NoStop}%
\bibitem [{\citenamefont {Jonay}\ \emph {et~al.}(2018)\citenamefont {Jonay}, \citenamefont {Huse},\ and\ \citenamefont {Nahum}}]{jonay2018coarse}%
  \BibitemOpen
  \bibfield  {author} {\bibinfo {author} {\bibfnamefont {C.}~\bibnamefont {Jonay}}, \bibinfo {author} {\bibfnamefont {D.~A.}\ \bibnamefont {Huse}},\ and\ \bibinfo {author} {\bibfnamefont {A.}~\bibnamefont {Nahum}},\ }\bibfield  {title} {\bibinfo {title} {Coarse-grained dynamics of operator and state entanglement},\ }\href@noop {} {\bibfield  {journal} {\bibinfo  {journal} {arXiv preprint arXiv:1803.00089}\ } (\bibinfo {year} {2018})}\BibitemShut {NoStop}%
\bibitem [{\citenamefont {Prosen}\ and\ \citenamefont {Pi{\v{z}}orn}(2007)}]{prosen2007operator}%
  \BibitemOpen
  \bibfield  {author} {\bibinfo {author} {\bibfnamefont {T.}~\bibnamefont {Prosen}}\ and\ \bibinfo {author} {\bibfnamefont {I.}~\bibnamefont {Pi{\v{z}}orn}},\ }\bibfield  {title} {\bibinfo {title} {Operator space entanglement entropy in a transverse ising chain},\ }\href@noop {} {\bibfield  {journal} {\bibinfo  {journal} {Physical Review A}\ }\textbf {\bibinfo {volume} {76}},\ \bibinfo {pages} {032316} (\bibinfo {year} {2007})}\BibitemShut {NoStop}%
\bibitem [{\citenamefont {Vidal}(2003)}]{vidal2003efficient}%
  \BibitemOpen
  \bibfield  {author} {\bibinfo {author} {\bibfnamefont {G.}~\bibnamefont {Vidal}},\ }\bibfield  {title} {\bibinfo {title} {Efficient classical simulation of slightly entangled quantum computations},\ }\href@noop {} {\bibfield  {journal} {\bibinfo  {journal} {Physical review letters}\ }\textbf {\bibinfo {volume} {91}},\ \bibinfo {pages} {147902} (\bibinfo {year} {2003})}\BibitemShut {NoStop}%
\bibitem [{\citenamefont {Hess}(2007)}]{hess2007heat}%
  \BibitemOpen
  \bibfield  {author} {\bibinfo {author} {\bibfnamefont {C.}~\bibnamefont {Hess}},\ }\bibfield  {title} {\bibinfo {title} {Heat conduction in low-dimensional quantum magnets},\ }\href@noop {} {\bibfield  {journal} {\bibinfo  {journal} {The European Physical Journal Special Topics}\ }\textbf {\bibinfo {volume} {151}},\ \bibinfo {pages} {73} (\bibinfo {year} {2007})}\BibitemShut {NoStop}%
\bibitem [{\citenamefont {Hirobe}\ \emph {et~al.}(2017)\citenamefont {Hirobe}, \citenamefont {Sato}, \citenamefont {Kawamata}, \citenamefont {Shiomi}, \citenamefont {Uchida}, \citenamefont {Iguchi}, \citenamefont {Koike}, \citenamefont {Maekawa},\ and\ \citenamefont {Saitoh}}]{hirobe2017one}%
  \BibitemOpen
  \bibfield  {author} {\bibinfo {author} {\bibfnamefont {D.}~\bibnamefont {Hirobe}}, \bibinfo {author} {\bibfnamefont {M.}~\bibnamefont {Sato}}, \bibinfo {author} {\bibfnamefont {T.}~\bibnamefont {Kawamata}}, \bibinfo {author} {\bibfnamefont {Y.}~\bibnamefont {Shiomi}}, \bibinfo {author} {\bibfnamefont {K.-i.}\ \bibnamefont {Uchida}}, \bibinfo {author} {\bibfnamefont {R.}~\bibnamefont {Iguchi}}, \bibinfo {author} {\bibfnamefont {Y.}~\bibnamefont {Koike}}, \bibinfo {author} {\bibfnamefont {S.}~\bibnamefont {Maekawa}},\ and\ \bibinfo {author} {\bibfnamefont {E.}~\bibnamefont {Saitoh}},\ }\bibfield  {title} {\bibinfo {title} {One-dimensional spinon spin currents},\ }\href@noop {} {\bibfield  {journal} {\bibinfo  {journal} {Nature Physics}\ }\textbf {\bibinfo {volume} {13}},\ \bibinfo {pages} {30} (\bibinfo {year} {2017})}\BibitemShut {NoStop}%
\bibitem [{\citenamefont {Bertini}\ \emph {et~al.}(2021)\citenamefont {Bertini}, \citenamefont {Heidrich-Meisner}, \citenamefont {Karrasch}, \citenamefont {Prosen}, \citenamefont {Steinigeweg},\ and\ \citenamefont {{\v{Z}}nidari{\v{c}}}}]{bertini2021finite}%
  \BibitemOpen
  \bibfield  {author} {\bibinfo {author} {\bibfnamefont {B.}~\bibnamefont {Bertini}}, \bibinfo {author} {\bibfnamefont {F.}~\bibnamefont {Heidrich-Meisner}}, \bibinfo {author} {\bibfnamefont {C.}~\bibnamefont {Karrasch}}, \bibinfo {author} {\bibfnamefont {T.}~\bibnamefont {Prosen}}, \bibinfo {author} {\bibfnamefont {R.}~\bibnamefont {Steinigeweg}},\ and\ \bibinfo {author} {\bibfnamefont {M.}~\bibnamefont {{\v{Z}}nidari{\v{c}}}},\ }\bibfield  {title} {\bibinfo {title} {Finite-temperature transport in one-dimensional quantum lattice models},\ }\href@noop {} {\bibfield  {journal} {\bibinfo  {journal} {Reviews of Modern Physics}\ }\textbf {\bibinfo {volume} {93}},\ \bibinfo {pages} {025003} (\bibinfo {year} {2021})}\BibitemShut {NoStop}%
\bibitem [{\citenamefont {Crosswhite}\ and\ \citenamefont {Bacon}(2008)}]{crosswhite2008finite}%
  \BibitemOpen
  \bibfield  {author} {\bibinfo {author} {\bibfnamefont {G.~M.}\ \bibnamefont {Crosswhite}}\ and\ \bibinfo {author} {\bibfnamefont {D.}~\bibnamefont {Bacon}},\ }\bibfield  {title} {\bibinfo {title} {Finite automata for caching in matrix product algorithms},\ }\href@noop {} {\bibfield  {journal} {\bibinfo  {journal} {Physical Review A}\ }\textbf {\bibinfo {volume} {78}},\ \bibinfo {pages} {012356} (\bibinfo {year} {2008})}\BibitemShut {NoStop}%
\bibitem [{\citenamefont {Huang}\ \emph {et~al.}(2013)\citenamefont {Huang}, \citenamefont {Karrasch},\ and\ \citenamefont {Moore}}]{huang2013scaling}%
  \BibitemOpen
  \bibfield  {author} {\bibinfo {author} {\bibfnamefont {Y.}~\bibnamefont {Huang}}, \bibinfo {author} {\bibfnamefont {C.}~\bibnamefont {Karrasch}},\ and\ \bibinfo {author} {\bibfnamefont {J.}~\bibnamefont {Moore}},\ }\bibfield  {title} {\bibinfo {title} {Scaling of electrical and thermal conductivities in an almost integrable chain},\ }\href@noop {} {\bibfield  {journal} {\bibinfo  {journal} {Physical review b}\ }\textbf {\bibinfo {volume} {88}},\ \bibinfo {pages} {115126} (\bibinfo {year} {2013})}\BibitemShut {NoStop}%
\bibitem [{\citenamefont {Hauschild}\ and\ \citenamefont {Pollmann}(2018)}]{hauschild2018efficient}%
  \BibitemOpen
  \bibfield  {author} {\bibinfo {author} {\bibfnamefont {J.}~\bibnamefont {Hauschild}}\ and\ \bibinfo {author} {\bibfnamefont {F.}~\bibnamefont {Pollmann}},\ }\bibfield  {title} {\bibinfo {title} {Efficient numerical simulations with tensor networks: Tensor network python (tenpy)},\ }\href@noop {} {\bibfield  {journal} {\bibinfo  {journal} {SciPost Physics Lecture Notes}\ ,\ \bibinfo {pages} {005}} (\bibinfo {year} {2018})}\BibitemShut {NoStop}%
\bibitem [{\citenamefont {Mukerjee}\ \emph {et~al.}(2006)\citenamefont {Mukerjee}, \citenamefont {Oganesyan},\ and\ \citenamefont {Huse}}]{Mukerjee_2006}%
  \BibitemOpen
  \bibfield  {author} {\bibinfo {author} {\bibfnamefont {S.}~\bibnamefont {Mukerjee}}, \bibinfo {author} {\bibfnamefont {V.}~\bibnamefont {Oganesyan}},\ and\ \bibinfo {author} {\bibfnamefont {D.}~\bibnamefont {Huse}},\ }\bibfield  {title} {\bibinfo {title} {Statistical theory of transport by strongly interacting lattice fermions},\ }\bibfield  {journal} {\bibinfo  {journal} {Physical Review B}\ }\textbf {\bibinfo {volume} {73}},\ \href {https://doi.org/10.1103/physrevb.73.035113} {10.1103/physrevb.73.035113} (\bibinfo {year} {2006})\BibitemShut {NoStop}%
\bibitem [{\citenamefont {Michailidis}\ \emph {et~al.}(2023)\citenamefont {Michailidis}, \citenamefont {Abanin},\ and\ \citenamefont {Delacrétaz}}]{michailidis2023corrections}%
  \BibitemOpen
  \bibfield  {author} {\bibinfo {author} {\bibfnamefont {A.~A.}\ \bibnamefont {Michailidis}}, \bibinfo {author} {\bibfnamefont {D.~A.}\ \bibnamefont {Abanin}},\ and\ \bibinfo {author} {\bibfnamefont {L.~V.}\ \bibnamefont {Delacrétaz}},\ }\href@noop {} {\bibinfo {title} {Corrections to diffusion in interacting quantum systems}} (\bibinfo {year} {2023}),\ \Eprint {https://arxiv.org/abs/2310.10564} {arXiv:2310.10564 [cond-mat.stat-mech]} \BibitemShut {NoStop}%
\bibitem [{\citenamefont {Bohrdt}\ \emph {et~al.}(2017)\citenamefont {Bohrdt}, \citenamefont {Mendl}, \citenamefont {Endres},\ and\ \citenamefont {Knap}}]{bohrdt2017scrambling}%
  \BibitemOpen
  \bibfield  {author} {\bibinfo {author} {\bibfnamefont {A.}~\bibnamefont {Bohrdt}}, \bibinfo {author} {\bibfnamefont {C.~B.}\ \bibnamefont {Mendl}}, \bibinfo {author} {\bibfnamefont {M.}~\bibnamefont {Endres}},\ and\ \bibinfo {author} {\bibfnamefont {M.}~\bibnamefont {Knap}},\ }\bibfield  {title} {\bibinfo {title} {Scrambling and thermalization in a diffusive quantum many-body system},\ }\href@noop {} {\bibfield  {journal} {\bibinfo  {journal} {New Journal of Physics}\ }\textbf {\bibinfo {volume} {19}},\ \bibinfo {pages} {063001} (\bibinfo {year} {2017})}\BibitemShut {NoStop}%
\bibitem [{\citenamefont {Chen-Lin}\ \emph {et~al.}(2019)\citenamefont {Chen-Lin}, \citenamefont {Delacr{\'{e} }taz},\ and\ \citenamefont {Hartnoll}}]{delacretaz_hartnoll19}%
  \BibitemOpen
  \bibfield  {author} {\bibinfo {author} {\bibfnamefont {X.}~\bibnamefont {Chen-Lin}}, \bibinfo {author} {\bibfnamefont {L.~V.}\ \bibnamefont {Delacr{\'{e} }taz}},\ and\ \bibinfo {author} {\bibfnamefont {S.~A.}\ \bibnamefont {Hartnoll}},\ }\bibfield  {title} {\bibinfo {title} {Theory of diffusive fluctuations},\ }\bibfield  {journal} {\bibinfo  {journal} {Physical Review Letters}\ }\textbf {\bibinfo {volume} {122}},\ \href {https://doi.org/10.1103/physrevlett.122.091602} {10.1103/physrevlett.122.091602} (\bibinfo {year} {2019})\BibitemShut {NoStop}%
\bibitem [{\citenamefont {Stoudenmire}\ and\ \citenamefont {White}(2012)}]{stoudenmire2012studying}%
  \BibitemOpen
  \bibfield  {author} {\bibinfo {author} {\bibfnamefont {E.~M.}\ \bibnamefont {Stoudenmire}}\ and\ \bibinfo {author} {\bibfnamefont {S.~R.}\ \bibnamefont {White}},\ }\bibfield  {title} {\bibinfo {title} {Studying two-dimensional systems with the density matrix renormalization group},\ }\href@noop {} {\bibfield  {journal} {\bibinfo  {journal} {Annu. Rev. Condens. Matter Phys.}\ }\textbf {\bibinfo {volume} {3}},\ \bibinfo {pages} {111} (\bibinfo {year} {2012})}\BibitemShut {NoStop}%
\bibitem [{\citenamefont {Brockt}\ \emph {et~al.}(2015)\citenamefont {Brockt}, \citenamefont {Dorfner}, \citenamefont {Vidmar}, \citenamefont {Heidrich-Meisner},\ and\ \citenamefont {Jeckelmann}}]{brockt2015matrix}%
  \BibitemOpen
  \bibfield  {author} {\bibinfo {author} {\bibfnamefont {C.}~\bibnamefont {Brockt}}, \bibinfo {author} {\bibfnamefont {F.}~\bibnamefont {Dorfner}}, \bibinfo {author} {\bibfnamefont {L.}~\bibnamefont {Vidmar}}, \bibinfo {author} {\bibfnamefont {F.}~\bibnamefont {Heidrich-Meisner}},\ and\ \bibinfo {author} {\bibfnamefont {E.}~\bibnamefont {Jeckelmann}},\ }\bibfield  {title} {\bibinfo {title} {Matrix-product-state method with a dynamical local basis optimization for bosonic systems out of equilibrium},\ }\href@noop {} {\bibfield  {journal} {\bibinfo  {journal} {Physical Review B}\ }\textbf {\bibinfo {volume} {92}},\ \bibinfo {pages} {241106} (\bibinfo {year} {2015})}\BibitemShut {NoStop}%
\end{thebibliography}%

\end{document}